\documentclass{aastex631}

\newcommand{\source}{VAST~J052348.6$-$712552}
\newcommand{\psr}{PSR~J0523$-$7125}

\newcommand{\sourceandy}{ASKAP~J173608.2$-$321635}

\received{xxx}
\revised{yyy}
\accepted{zzz}

\submitjournal{ApJ}

\shorttitle{Discovery of \psr\ with ASKAP}
\shortauthors{Wang et al.}
\graphicspath{{./}{figures/}}

\begin{document}

\title{Discovery of \psr\ as a Circularly Polarized Variable Radio Source in the Large Magellanic Cloud}

\correspondingauthor{Yuanming Wang}
\email{ywan3191@uni.sydney.edu.au, tara.murphy@sydney.edu.au}

\author[0000-0003-0203-1196]{Yuanming Wang}
\affiliation{Sydney Institute for Astronomy, School of Physics, The University of Sydney, Sydney, New South Wales 2006, Australia}
\affiliation{Australia Telescope National Facility, CSIRO, Space and Astronomy, PO Box 76, Epping, NSW 1710, Australia}
\affiliation{ARC Centre of Excellence for Gravitational Wave Discovery (OzGrav), Hawthorn, Victoria, Australia}

\author[0000-0002-2686-438X]{Tara Murphy}
\affiliation{Sydney Institute for Astronomy, School of Physics, The University of Sydney, Sydney, New South Wales 2006, Australia}
\affiliation{ARC Centre of Excellence for Gravitational Wave Discovery (OzGrav), Hawthorn, Victoria, Australia}

\author[0000-0001-6295-2881]{David~L.~Kaplan}
\affiliation{Center for Gravitation, Cosmology, and Astrophysics, Department of Physics, University of Wisconsin-Milwaukee, P.O. Box 413, Milwaukee, WI 53201, USA}

\author[0000-0002-8768-266X]{Teresa Klinner-Teo}
\affiliation{Sydney Institute for Astronomy, School of Physics, The University of Sydney, Sydney, New South Wales 2006, Australia}

\author[0000-0001-6762-2638]{Alessandro Ridolfi}
\affiliation{INAF -- Osservatorio Astronomico di Cagliari, Via della Scienza 5, I-09047 Selargius (CA), Italy}
\affiliation{Max-Planck-Institut f\"{u}r Radioastronomie, Auf dem H\"{u}gel 69, D-53121 Bonn, Germany}


\author[0000-0003-3294-3081]{Matthew Bailes}
\affiliation{ARC Centre of Excellence for Gravitational Wave Discovery (OzGrav), Hawthorn, Victoria, Australia}
\affiliation{Centre for Astrophysics and Supercomputing, Swinburne University of Technology, Hawthorn, Victoria, Australia}

\author[0000-0002-2578-0360]{Fronefield Crawford}
\affiliation{Department of Physics and Astronomy, Franklin and Marshall College, Lancaster, PA 17604-3003, USA}

\author[0000-0002-9618-2499]{Shi Dai}
\affiliation{School of Science, Western Sydney University, Locked Bag 1797, Penrith South DC, NSW 2751, Australia}

\author[0000-0003-0699-7019]{Dougal Dobie}
\affiliation{ARC Centre of Excellence for Gravitational Wave Discovery (OzGrav), Hawthorn, Victoria, Australia}
\affiliation{Centre for Astrophysics and Supercomputing, Swinburne University of Technology, Hawthorn, Victoria, Australia}

\author[0000-0002-3382-9558]{B.~M.~Gaensler}
\affiliation{Dunlap Institute for Astronomy and Astrophysics, University of Toronto, 50 St. George St., Toronto, ON M5S 3H4, Canada}
\affiliation{David A. Dunlap Department of Astronomy and Astrophysics, University of Toronto, 50 St. George St., Toronto, ON M5S 3H4, Canada}

\author[0000-0002-6558-1681]{Vanessa Graber}
\affiliation{Institute of Space Sciences (ICE, CSIC), Campus UAB, Carrer de Can Magrans s/n, 08193, Barcelona, Spain}

\author[0000-0001-6864-5057]{Ian Heywood}
\affiliation{Astrophysics, Department of Physics, University of Oxford, Keble Road, Oxford, OX1 3RH, UK}
\affiliation{Centre for Radio Astronomy Techniques and Technologies, Department of Physics and Electronics, Rhodes University, PO Box 94, Makhanda, 6140, South Africa}
\affiliation{South African Radio Astronomy Observatory, 2 Fir Street, Black River Park, Observatory, Cape Town, 7925, South Africa}

\author[0000-0002-9994-1593]{Emil Lenc}
\affiliation{Australia Telescope National Facility, CSIRO, Space and Astronomy, PO Box 76, Epping, NSW 1710, Australia}

\author[0000-0003-1301-966X]{Duncan R.\ Lorimer}
\affiliation{Department of Physics and Astronomy, West Virginia University, Morgantown, WV 26501, USA}
\affiliation{Center for Gravitational Waves and Cosmology, West Virginia University, Chestnut Ridge Research Building, Morgantown, WV 26505, USA}

\author[0000-0001-7697-7422]{Maura A.\ McLaughlin}
\affiliation{Department of Physics and Astronomy, West Virginia University, Morgantown, WV 26501, USA}
\affiliation{Center for Gravitational Waves and Cosmology, West Virginia University, Chestnut Ridge Research Building, Morgantown, WV 26505, USA}

\author[0000-0003-4609-2791]{Andrew O'Brien}
\affiliation{Center for Gravitation, Cosmology, and Astrophysics, Department of Physics, University of Wisconsin-Milwaukee, P.O. Box 413, Milwaukee, WI 53201, USA}

\author[0000-0003-3860-5825]{Sergio Pintaldi}
\affiliation{Sydney Informatics Hub, The University of Sydney, NSW 2008, Australia}

\author[0000-0003-1575-5249]{Joshua Pritchard}
\affiliation{Sydney Institute for Astronomy, School of Physics, The University of Sydney, Sydney, New South Wales 2006, Australia}
\affiliation{Australia Telescope National Facility, CSIRO, Space and Astronomy, PO Box 76, Epping, NSW 1710, Australia}
\affiliation{ARC Centre of Excellence for Gravitational Wave Discovery (OzGrav), Hawthorn, Victoria, Australia}

\author[0000-0003-2177-6388]{Nanda Rea}
\affiliation{Institute of Space Sciences (ICE, CSIC), Campus UAB, Carrer de Can Magrans s/n, 08193, Barcelona, Spain}
\affiliation{Institut d'Estudis Espacials de Catalunya (IEEC), Carrer Gran Capit\`a 2–4, 08034 Barcelona, Spain}

\author[0000-0002-3017-092X]{Joshua P.\ Ridley}
\affiliation{School of Engineering, Murray State University, Murray, KY 42071, USA}

\author[0000-0003-2781-9107]{Michele Ronchi}
\affiliation{Institute of Space Sciences (ICE, CSIC), Campus UAB, Carrer de Can Magrans s/n, 08193, Barcelona, Spain}

\author[0000-0002-7285-6348]{Ryan M. Shannon}
\affiliation{ARC Centre of Excellence for Gravitational Wave Discovery (OzGrav), Hawthorn, Victoria, Australia}
\affiliation{Centre for Astrophysics and Supercomputing, Swinburne University of Technology, Hawthorn, Victoria, Australia}

\author[0000-0001-6682-916X]{Gregory R. Sivakoff}
\affiliation{Department of Physics, University of Alberta, CCIS 4-181, Edmonton, AB T6G 2E1, Canada}

\author[0000-0001-8026-5903]{Adam Stewart}
\affiliation{Sydney Institute for Astronomy, School of Physics, The University of Sydney, Sydney, New South Wales 2006, Australia}

\author[0000-0002-2066-9823]{Ziteng Wang}
\affiliation{Sydney Institute for Astronomy, School of Physics, The University of Sydney, Sydney, New South Wales 2006, Australia}
\affiliation{Australia Telescope National Facility, CSIRO, Space and Astronomy, PO Box 76, Epping, NSW 1710, Australia}
\affiliation{ARC Centre of Excellence for Gravitational Wave Discovery (OzGrav), Hawthorn, Victoria, Australia}

\author[0000-0002-9583-2947]{Andrew Zic}
\affiliation{Department of Physics and Astronomy, and Research Centre in Astronomy, Astrophysics and Astrophotonics, Macquarie University, NSW 2109, Australia}
\affiliation{Australia Telescope National Facility, CSIRO, Space and Astronomy, PO Box 76, Epping, NSW 1710, Australia}

\begin{abstract}

We report the discovery of a highly circularly polarized,  variable, steep-spectrum pulsar in the Australian Square Kilometre Array Pathfinder (ASKAP) Variables and Slow Transients (VAST) survey. 
The pulsar is located about $1^\circ$ from \added{the center of} the Large Magellanic Cloud, and has a significant fractional circular polarization of $\sim$20\%. 
We discovered pulsations with a period of 322.5\,ms, dispersion measure (DM) of 157.5\,pc\,cm$^{-3}$, and rotation measure (RM) of $+456$\,rad\,m$^{-2}$ using observations from the MeerKAT and the Parkes telescopes.  
This DM firmly places the source, \psr, in the Large Magellanic Cloud (LMC). 
This RM is extreme compared to other pulsars in the LMC (more than twice that of the largest previously reported one). 
The average flux density of $\sim$1\,mJy at 1400\,MHz and $\sim$25\,mJy at 400\,MHz places it among the most luminous radio pulsars known.
It likely evaded previous discovery because of its very steep radio spectrum (spectral index $\alpha \approx -3$, where $S_\nu \propto \nu^\alpha$) and broad pulse profile (duty cycle $\gtrsim35\%$).  
We discuss implications for searches for unusual radio sources in continuum images, 
as well as extragalactic pulsars in the Magellanic Clouds and beyond.
Our result highlighted the possibility of identifying pulsars, especially extreme pulsars, from radio continuum images. 
Future large-scale radio surveys will give us an unprecedented opportunity to discover more pulsars and potentially the most distant pulsars beyond the Magellanic Clouds. 
\end{abstract}


\section{Introduction} 
\label{sec:introduction}


Since the discovery of the first pulsar~\citep{1968Natur.217..709H}, a lot of effort has been devoted to developing efficient and sensitive search algorithms (e.g., \citealt{2001PhDT.......123R, Ransom2003}). 
The traditional radio pulsar search procedures focus on the Fourier domain or time domain to identify periodic signals (see Chapter~6 in \citealt{2012hpa..book.....L}). 
While this has been fruitful and has discovered more than $\sim$3000 pulsars~\citep{2005AJ....129.1993M}, abnormal pulsars such as short orbital period binary systems or strongly scattered objects are more difficult to detect. 
Indeed, some algorithms have been developed (e.g., for acceleration and ``jerk'' searches; \citealt{2001PhDT.......123R, Andersen2018}) to explore a broader parameter space, but they are computationally expensive for untargeted sky surveys. 

Continuum images have long been considered as an effective way to select pulsar candidates based on their steep spectral indices \citep[e.g.,][]{1982Natur.300..615B, 1985PASA....6..174D, 1987ApJ...319L.103S, 2016MNRAS.461.1062F, Bhakta2017, 2018ApJ...864...16M}. 
Since continuum images are sensitive to pulsar emission regardless of period, scattering or orbital modulation, in principle they can allow us to find extreme pulsars that have not been detected in traditional pulsar surveys. 
In fact, the first known millisecond pulsar was initially noticed through the unusual, steep spectral properties, and scintillation of the continuum source, and only then was confirmed by a targeted pulsar search \citep{1982Natur.300..615B}. 

Pulsars are known to be compact enough to exhibit variability due to interstellar scintillation, caused by irregularities in the turbulent interstellar medium \citep{1990ARA&A..28..561R}. 
\citet{2016MNRAS.462.3115D} proposed a new detection technique based on this; using diffractive interstellar scintillation in variance images. 
The scintillation behavior, plus potential intrinsic fluctuations (e.g., intermittent behavior; \citealt{2009ASSL..357...67L}), will cause strong flux density variations, and therefore pulsars can sometimes be detected in general radio variability surveys \citep[e.g.,][]{2016MNRAS.461..908B, 2021PASA...38...54M}. 

Another approach for identifying pulsars in continuum surveys is through circularly polarized emission \citep{Gaensler1998, 2019ApJ...884...96K}. 
Only a few types of sources are known to be more than a few percent circularly polarized; they are usually pulsars \citep[e.g.,][]{2015MNRAS.449.3223D,2018MNRAS.478.2835L} or stellar objects \citep[e.g.,][]{2021MNRAS.502.5438P}. 
Thus, highly circularly polarized sources that lack a deep optical/infrared counterpart are strong pulsar candidates.
To-date, there have only been two large-scale circular polarization surveys. 
\citet{2018MNRAS.478.2835L} conducted the first all-sky circular polarization survey using the Murchison Widefield Array (MWA) at 200\,MHz, and detected 14 known pulsars in the untargeted survey. 
After that, \citet{2021MNRAS.502.5438P} performed a circular polarization survey for radio stars with the Australian Square Kilometre Array Pathfinder (ASKAP; \citealt{Hotan2021}) at 888 MHz, and identified 33 known pulsars. 

We use the data from the ASKAP Phase I Pilot survey for Variables and Slow Transients (VAST-P1; \citealt{2021PASA...38...54M}) processed with the VAST pipeline \citep{2021arXiv210105898P} to search for variable and transient sources in the two ASKAP fields covering the Magellanic Clouds. 
Here we report the discovery of a highly-variable, circularly-polarized, steep-spectrum source \source\ in this continuum survey. 
After targeted MeerKAT and Parkes follow-up observations, we identified it as a new pulsar, \psr\ located in the Large Magellanic Cloud (LMC). 
The extragalactic distance makes it one of the most luminous pulsars known at both 400\,MHz ($L_\mathrm{400}\approx6.3\times10^4$\,mJy\,kpc$^2$) and 1400\,MHz ($L_\mathrm{1400}\approx2.5\times10^3$\,mJy\,kpc$^2$). 
We present our observations and results in Section~\ref{sec:observations}. 
In Section~\ref{sec:discussion}, we discuss the nature of the pulsar and prospects for identifying future pulsars through continuum imaging surveys.

\section{Observations and Results} 
\label{sec:observations}

\subsection{ASKAP discovery} 
\label{subsec:askap_discovery}

We observed two 30~deg$^2$ fields (field names VAST 0530$-$68A and VAST 0127$-$73A) centered on the Magellanic Clouds, six times between 2019~August and 2020~January. 
Each observation had an integration time of 12\,minutes, achieving a typical rms sensitivity of 0.25\,mJy\,beam$^{-1}$ \added{and angular resolution of 12$''$} at a central frequency of 888\,MHz with a bandwidth of 288\,MHz. 
VAST-P1 incorporated data from the Rapid ASKAP Continuum Survey \citep[RACS;][]{2020PASA...37...48M} as the first epoch, which has the same observing frequency, but a longer integration time of 15\,minutes, resulting in a rms sensitivity of 0.20\,mJy\,beam$^{-1}$ for regions near the Magellanic Clouds. 
Details of data reduction for these surveys are given in \citet{2020PASA...37...48M} and \citet{2021PASA...38...54M}. 

\begin{figure*}
    \centering
	\includegraphics[width=\textwidth]{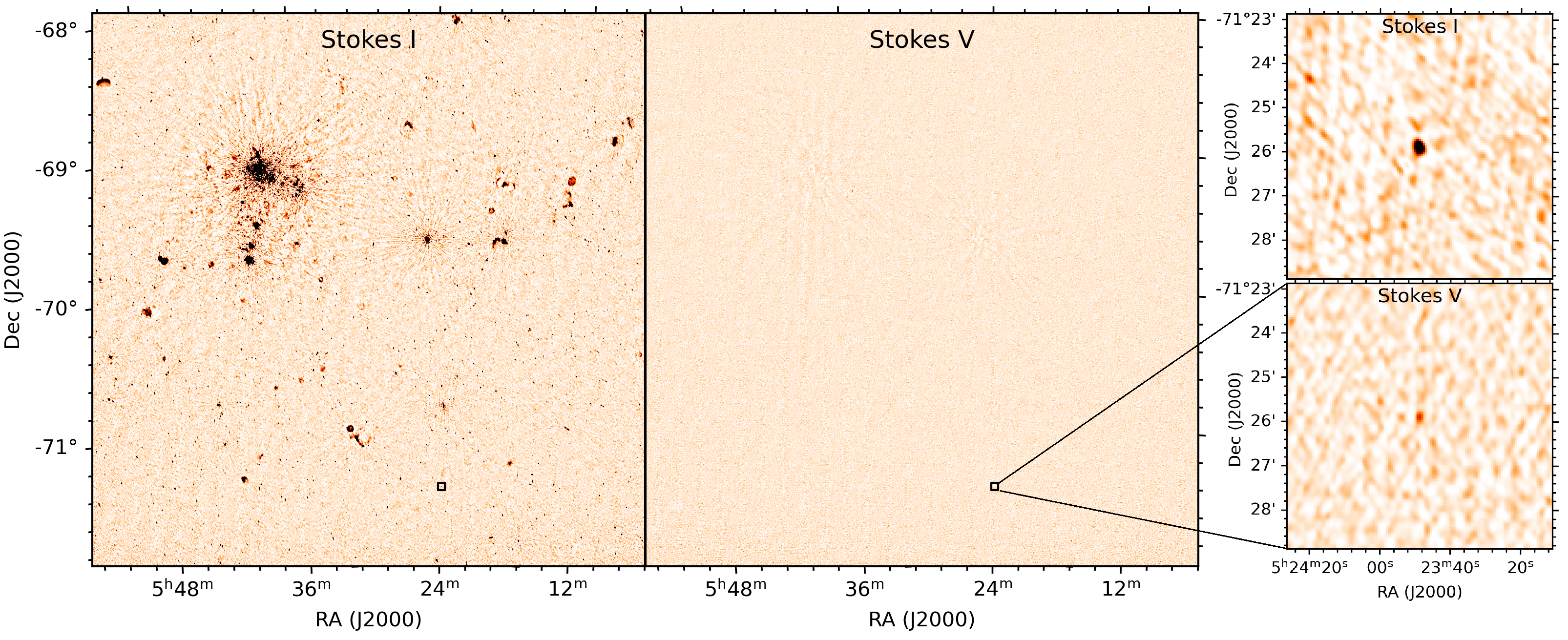}
    \caption{The ASKAP 888~MHz image of the LMC field. We show the total intensity (Stokes I) image on the left, and the circular polarization (Stokes V) image in the center. On the right we show the zoomed images in total intensity (top) and circular polarization (bottom) centered at the position of \source. }
    \label{fig:vast_cutout}
\end{figure*}

\begin{figure*}
    \centering
	\includegraphics[width=0.8\textwidth]{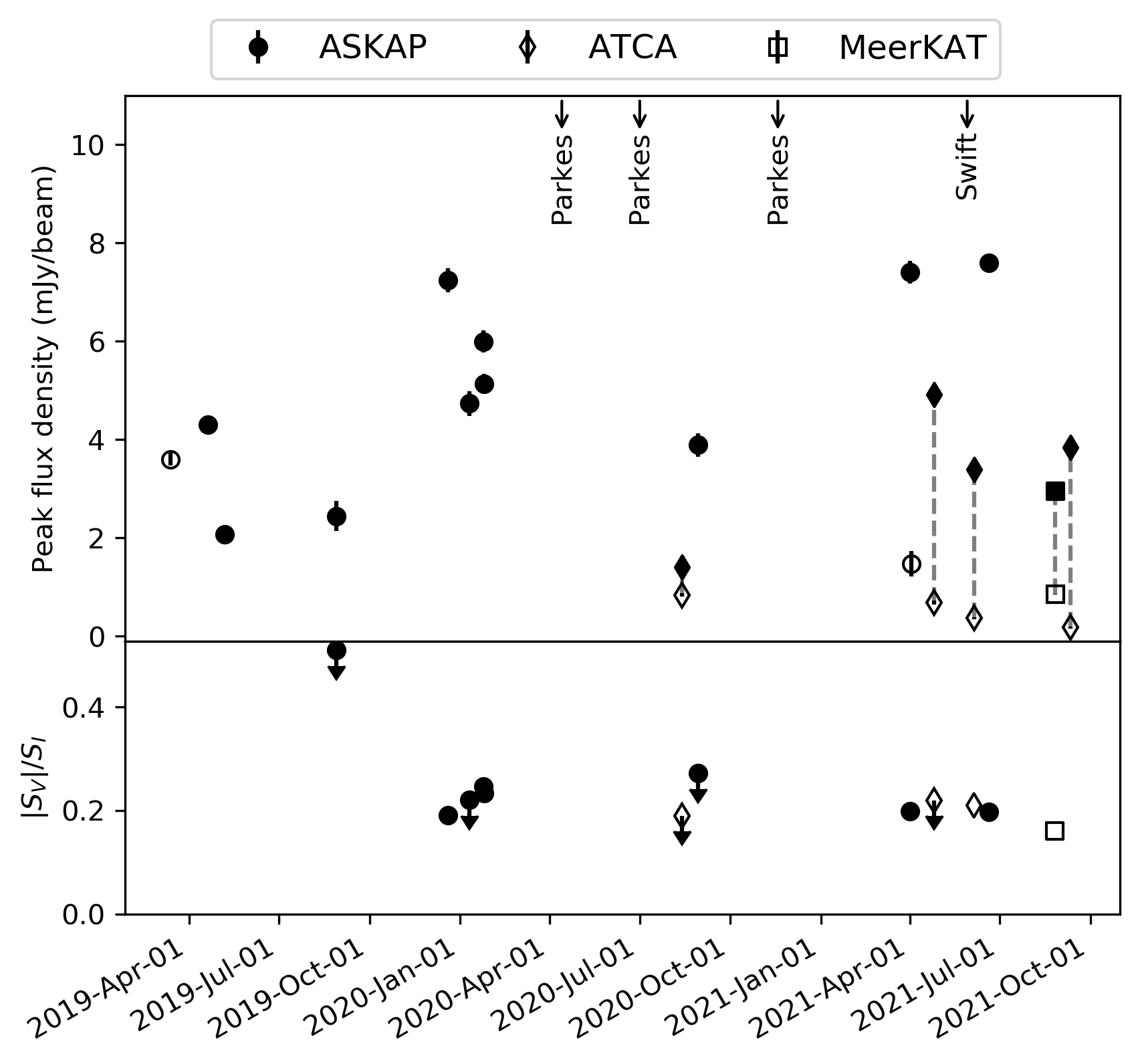}
    \caption{The radio lightcurve of \source, including total intensity (Stokes I) and fractional circular-polarization ($|S_V|/S_I$) measured at every available epoch. The closed markers represent the flux density at 888\,MHz, and the open markers represent the flux measurement at different frequencies: \replaced{(1337\,MHz for MeerKAT, 1367.5\,MHz for ASKAP mid-band, and 2100\,MHz for ATCA)}{1284\,MHz for MeerKAT (open squares), 1420\,MHz and 1367.5\,MHz for ASKAP mid-band at 2019~Mar~13 and 2021~Apr~2, respectively (open circles), and 2100\,MHz for ATCA (open diamonds)}. We connect the original flux density (open marker; measured at higher frequency with ATCA and MeerKAT) and scaled flux density (closed marker; scaled to 888\,MHz based on fitted spectral index at that epoch) by a gray dashed line. The upper limits for $|S_V|/S_I$ are from Stokes V measurements at the $5\sigma$ rms level. The black arrows on the top indicate the date of multiwavelength observations. }
    \label{fig:lightcurve}
\end{figure*}

We conducted a search for highly variable sources in the two fields using the VAST transient detection pipeline \citep[][]{2021arXiv210105898P, 2021PASA...38...54M}. We selected candidates with a variability index $\mathcal{V}>1.0\sigma_\mathcal{V}$ (equivalent to the fractional variability used by other surveys) and reduced chi-square $\eta>2.0\sigma_\eta$ (equating to values of $\mathcal{V}>0.295$ and $\eta>6.479$), where $\sigma$ is the standard deviation measured by fitting a Gaussian function to the distributions of both metrics in logarithmic space (following the calculations in \citealt{2019A&C....27..111R}). 
We identified $\sim$27 candidates as highly variable, compact sources. 
Two of them were detected with strong circular polarization, and \source\ was the only one lacking a clear multiwavelength association/identification (the other one is associated with a stellar object). 
The fractional circular polarization of \source\ ($f_\mathrm{p}=|S_V|/S_I$) is about 15--30\% (see the cutout in Figure~\ref{fig:vast_cutout}). 
No linear polarization was detected for this source in VAST-P1, placing a $3\sigma$ upper limit of $\sim$0.9\,mJy\,beam$^{-1}$ ($\lesssim10\%$ fractional polarization) on the total linear polarisation ($P=\sqrt{Q^2+U^2}$). 
The radio observations are summarised in Table~\ref{tab:observations} and shown in Figure~\ref{fig:lightcurve}. 
Other variable sources and the general survey results will be published in a subsequent paper. 

\subsection{MeerKAT and Parkes observations}
\label{subsec:parkes_meerkat}

\begin{figure*}
    \centering
	\includegraphics[width=\textwidth]{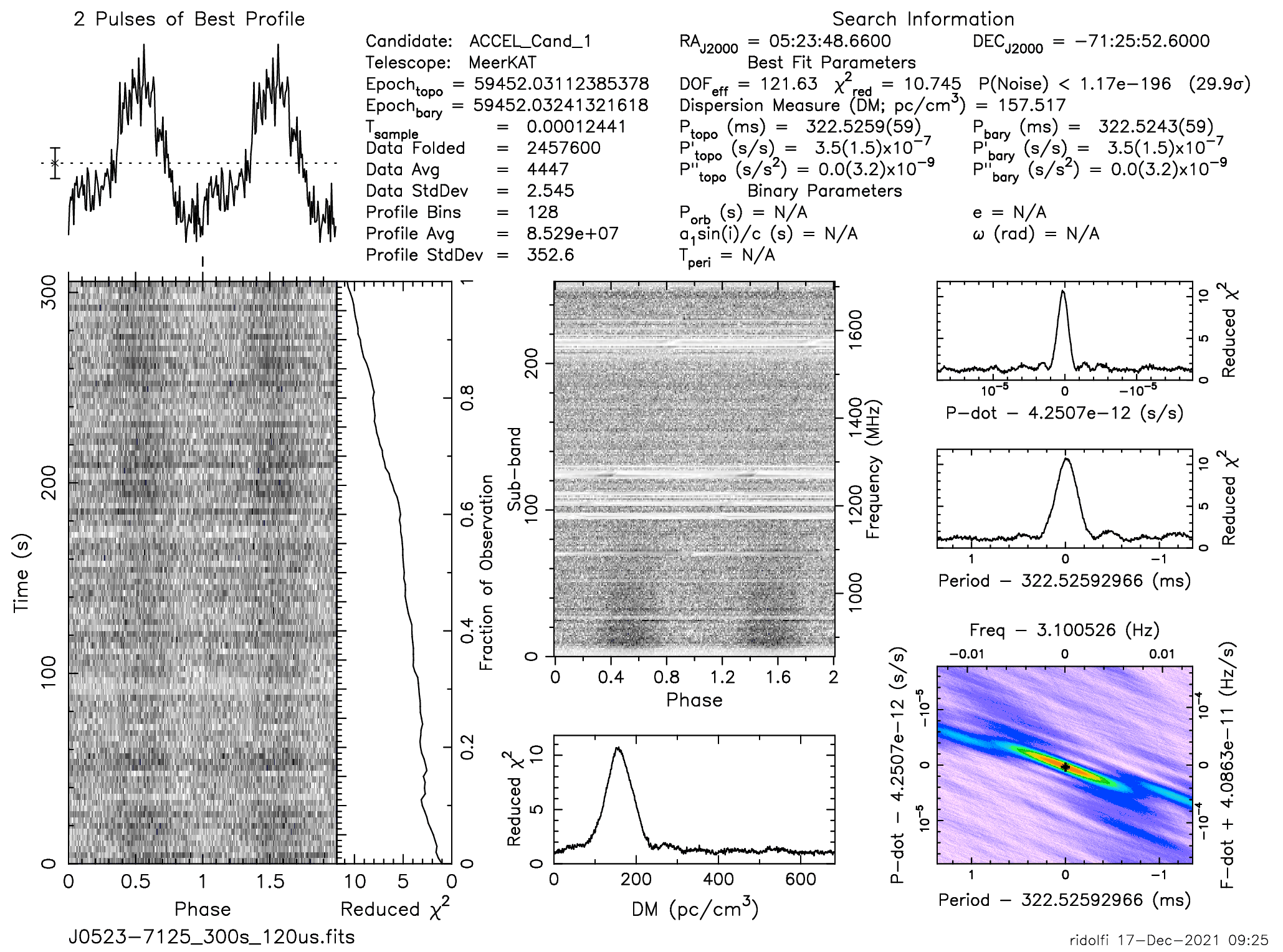}
    \caption{The initial discovery of \psr\ from MeerKAT. Note this is from the first 300\,s of the full 2.5\,h MeerKAT observation. The plot was generated using \textsc{presto}. }
    \label{fig:pulsar}
\end{figure*}

\begin{figure*}
    \centering
	\includegraphics[width=0.49\textwidth]{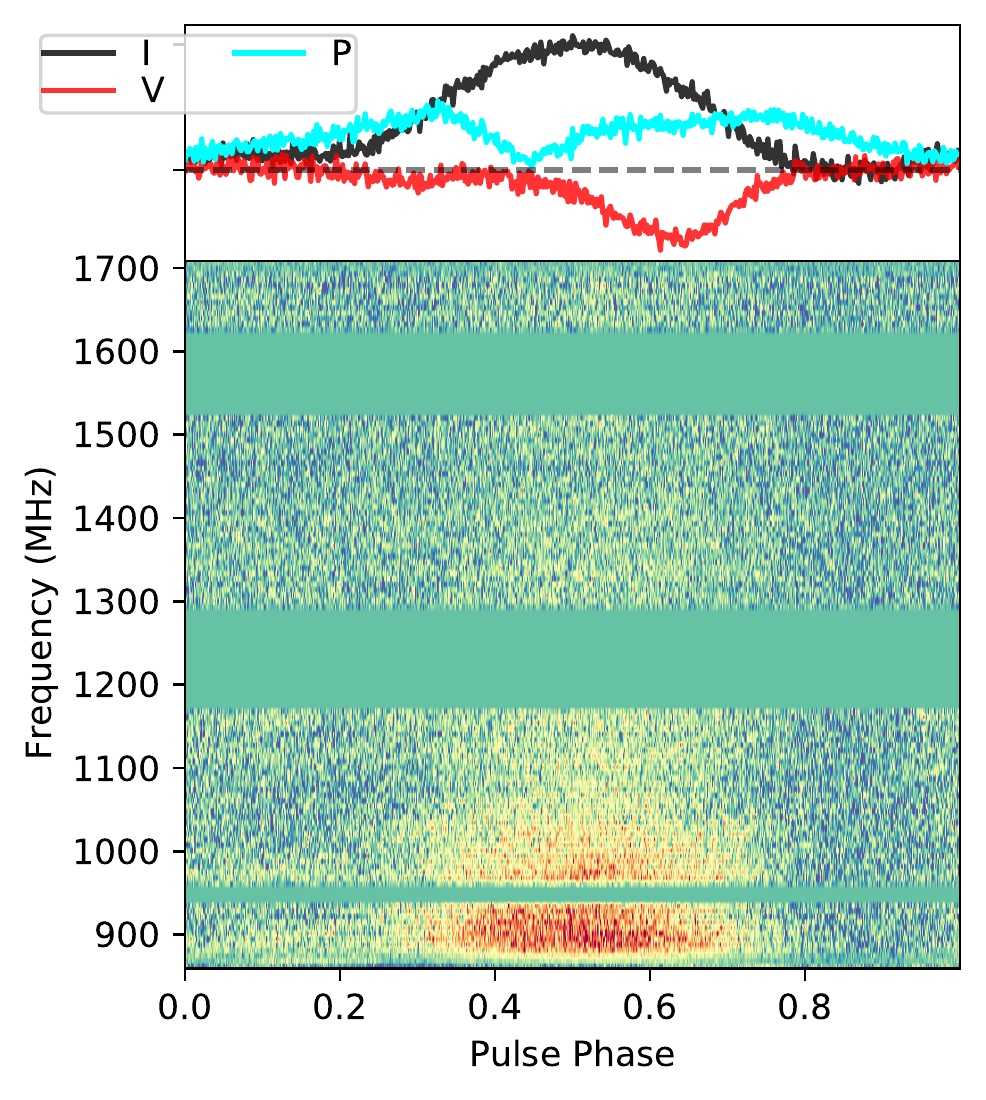}
	\includegraphics[width=0.49\textwidth]{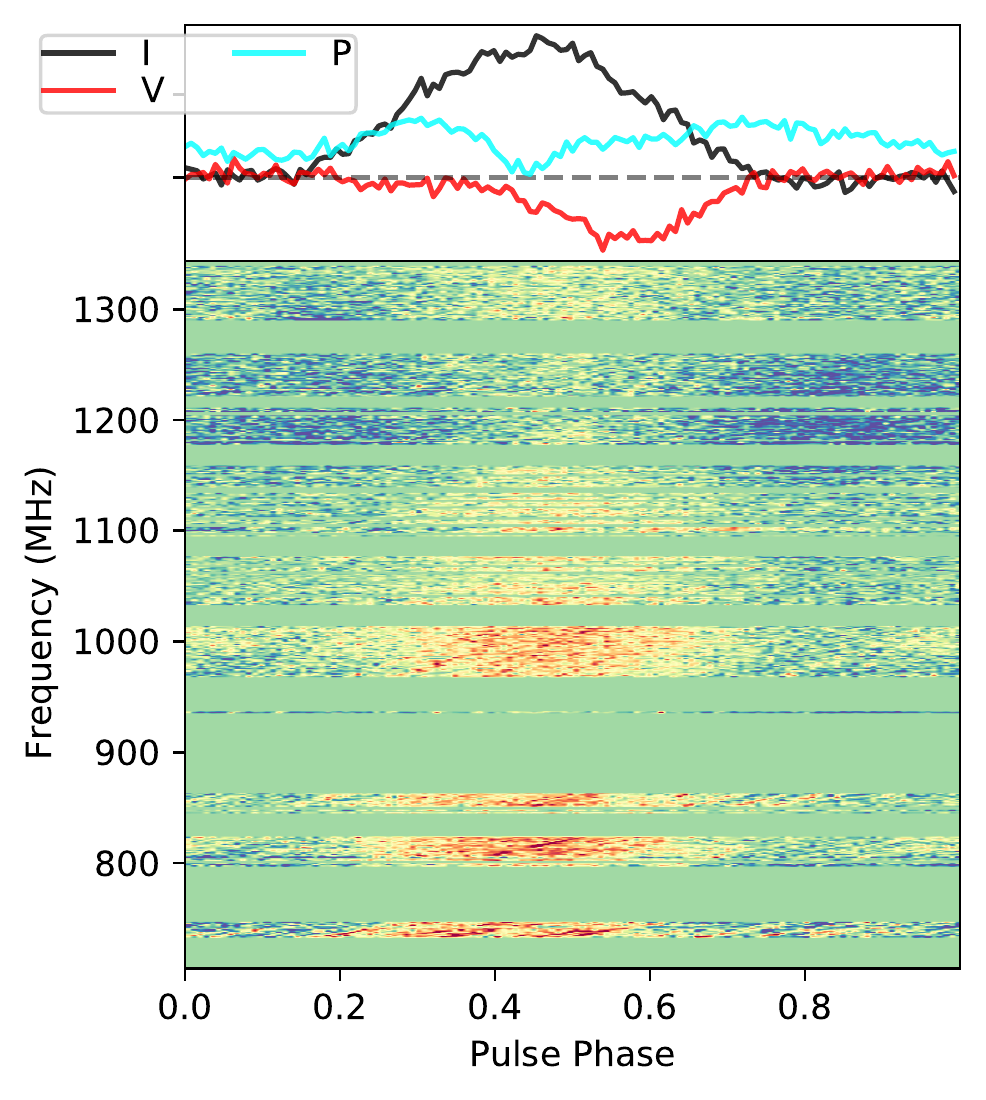}
    \caption{Folded pulse profiles of \psr, from the MeerKAT observation (left) and the Parkes observation (right). We show the pulse intensity as a function of frequency over the full bandpass for MeerKAT (lower left) and only the lower UWL bandpass (lower right). The integrated pulse profiles (top) are summed over frequencies $\leq1100$\,MHz, where the signal-to-noise ratio is best. We also show the polarized intensity profiles for both MeerKAT and Parkes observation (with RM correction). Note the linear polarization ($P$) is (unphysically) higher than total intensity ($I$) at some pulse phases, possibly due to a combination of RFI issues and `off-pulse baseline' subtraction uncertainty. }
    
    \label{fig:pulsar_profile}
\end{figure*}

We observed \source\ with the MeerKAT radio telescope (pulsar search mode and continuum imaging mode simultaneously) for 2.5\,h at a central frequency of 1284\,MHz (bandwidth of 856\,MHz) on 2021~August~25. 
We searched the MeerKAT (pulsar mode) data for pulsar candidates using the standard Fourier domain search procedure. 
This was done using \textsc{pulsar\_miner}\footnote{\url{https://github.com/alex88ridolfi/PULSAR\_MINER}} (see \citealt{Ridolfi+2021} for more details), an automated pipeline based on the \textsc{presto}\footnote{\url{https://github.com/scottransom/presto}} pulsar searching package \citep{Ransom2011}. 
After cleaning the observing band by removing the frequency channels affected by strong radio frequency interference (RFI), we generated de-dispersed time series, correcting for the interstellar dispersion with dispersion measure (DM) trial values in the range $2.0\textrm{--}300$~pc\,cm$^{-3}$, with steps of 0.05~pc\,cm$^{-3}$. 
Each time series was Fourier transformed and the resulting power spectrum searched for prominent periodicities using \textsc{presto}'s \texttt{accelsearch} routine. 
The latter is sensitive to both isolated and binary pulsars, by accounting for possible Doppler shifts of the pulsar spin frequency in the Fourier domain due to orbital motion (see \citealt{Ransom+2002} for a detailed discussion of the acceleration search technique). 
For our search, we allowed a maximum drift of up to 200 Fourier bins, using the \texttt{-zmax 200} option of \texttt{accelsearch}.
We identified a strong pulsar candidate, \psr, with a period of 322.5\,ms at a DM of 157.5\,pc\,cm$^{-3}$. 
No significant acceleration was detected within the 2.5 hours of the observation, indicating that the candidate pulsar is likely isolated, although further timing is ongoing.
Figure~\ref{fig:pulsar} shows the initial discovery plot of the pulsar. 
It has a very steep spectrum, consistent with the measurement in continuum imaging (see described below) and a wide pulse profile with duty cycle $\sim$35\% (using the pulse width at 50 percent maximum, $W_{50}$). 
Note the duty cycle may be underestimated as it is difficult to identify the off pulse baseline for a (wide pulsed) pulsar like this. 
We also measured a $\textrm{RM}=+456\pm6$\,rad\,m$^{-2}$ using the \texttt{rmfit} tool in \textsc{psrchive} \citep{2004PASA...21..302H}, and found a significant linear polarization ($\gtrsim50\%$) after RM correction (see pulse profiles in Figure~\ref{fig:pulsar_profile}). 
Further flux calibration is necessary to measure precise polarization properties.  
This pulsar has not been recorded in any known pulsar catalog \citep{2005AJ....129.1993M} or in the Pulsar Survey Scraper\footnote{\url{https://pulsar.cgca-hub.org}.}, which records newly-discovered pulsars prior to publication.

We also processed the MeerKAT (simultaneous) continuum data using \textsc{oxkat}~\citep{2020ascl.soft09003H}, which uses: \textsc{casa} for basic flagging, cross calibration, and splitting out measurement sets; \textsc{tricolour}\footnote{\url{https://github.com/ska-sa/tricolour}} for further flagging; \textsc{cubical}~\citep{2018MNRAS.478.2399K} for self-calibration; and \textsc{wsclean}~\citep{2014MNRAS.444..606O} for continuum imaging. 
We used PMN~J0408$-$6545 for bandpass and flux calibration, and PMN~J0420$-$6223 for phase calibration. 
We detected \source\ with a total flux density (Stokes $I$) of $860\pm7\,\mu$Jy\,beam$^{-1}$ (at $\gtrsim100\sigma$ confidence level), and circularly-polarized flux density (Stokes $|V|$) of $137\pm5\,\mu$Jy\,beam$^{-1}$. 
\added{The source is unsolved in the image, which has an angular resolution of $7.5''$. }
In contrast to the pulsar mode result, no linear polarization was detected above 3$\sigma$ threshold, and a Rotation Measure (RM) synthesis found no significant polarized intensity in $|\mathrm{RM}|<1200$\,rad\,m$^{-2}$. 
We will discuss this inconsistency in Section~\ref{subsec:pulsar_properties}. 
We obtained the (Stokes I) spectrum from 8 sub-band images and noticed an unusual upturn at higher frequency (see Figure~\ref{fig:spectral}). 
We fit a smoothly broken power-law\footnote{Following \url{https://docs.astropy.org/en/stable/api/astropy.modeling.powerlaws.SmoothlyBrokenPowerLaw1D.html}} to the spectrum, and found an upturn at $1.36\pm0.02$\,GHz with a spectral index $\alpha=-4.40\pm0.11$ in the lower frequency range and $\alpha=+2.34\pm0.51$ in the upper frequency range. 
We also obtained lightcurves (with 10\,min, 2\,min, and 8\,sec resolution, respectively) within the 2.5\,h observation, and found no significant variability on those timescales. 

We observed \source\ with the 64-m Parkes telescope on 2020~April~13, 2020~July~1 and 2020~November~18 using the pulsar searching mode with the Ultra-Wideband Low (UWL) receiver (which provides simultaneous frequency coverage from 704 to 4032\,MHz; \citealt{2020PASA...37...12H}). 
We used {\sc presto} \citep{2001PhDT.......123R} to perform a standard pulsar search of DMs spanning 0$-$300\,pc\,cm$^{-3}$, plus an acceleration search up to $\sim$200\,m\,s$^{-2}$. 
Our initial search of the Parkes observations found some candidate pulsars, but none were convincing. 
After detecting the pulsar with MeerKAT we reinspected our data and found a pulsar candidate with a similar period and DM. 
It had not been identified as a convincing candidate in our initial search since the signal is very weak (near our searching threshold SNR of $\sim$8), possibly due to the wide pulse profile and the steep spectrum (so it is only bright at low frequencies). 
We then re-analyzed the Parkes observations using the low frequency data only, and successfully detected this pulsar at frequencies $\lesssim2$\,GHz. 
Our Parkes data also measured a similar $\textrm{RM}=+457.6\pm0.1$\,rad\,m$^{-2}$ using \textsc{rmfit}. 

We note that this sky region was searched as part of the survey conducted by \citet{Ridley2013} using the Parkes telescope, who claimed a flux density limit of $0.05\,$mJy at 1400\,MHz, well below the flux density of \psr. 
However, we were able to identify a weak candidate in their data as well, with parameters consistent with \psr. 
Note that the \citet{Ridley2013} limit assumed a 5\% pulse width (compared to 35\% for \psr) and assumed that sources were at the centers of the beams of the Parkes Multi-Beam Receiver.
Accounting for our pulse width increases the limit by $\sim3$, while accounting for the position (as the pulsar is close to the half-power point of one of the beams) increases it by a further factor of $\sim2$. 
This gives a limiting flux density of about 0.3\,mJy, still less than what we observe, but the remaining differences may be due to a combination of scintillation, lost bandwidth due to radio frequency interference, and similar effects.

\subsection{ATCA and archival radio observations}
\label{subsec:atca_archival_radio}

\begin{figure*}
    \centering
    \includegraphics[width=0.49\textwidth]{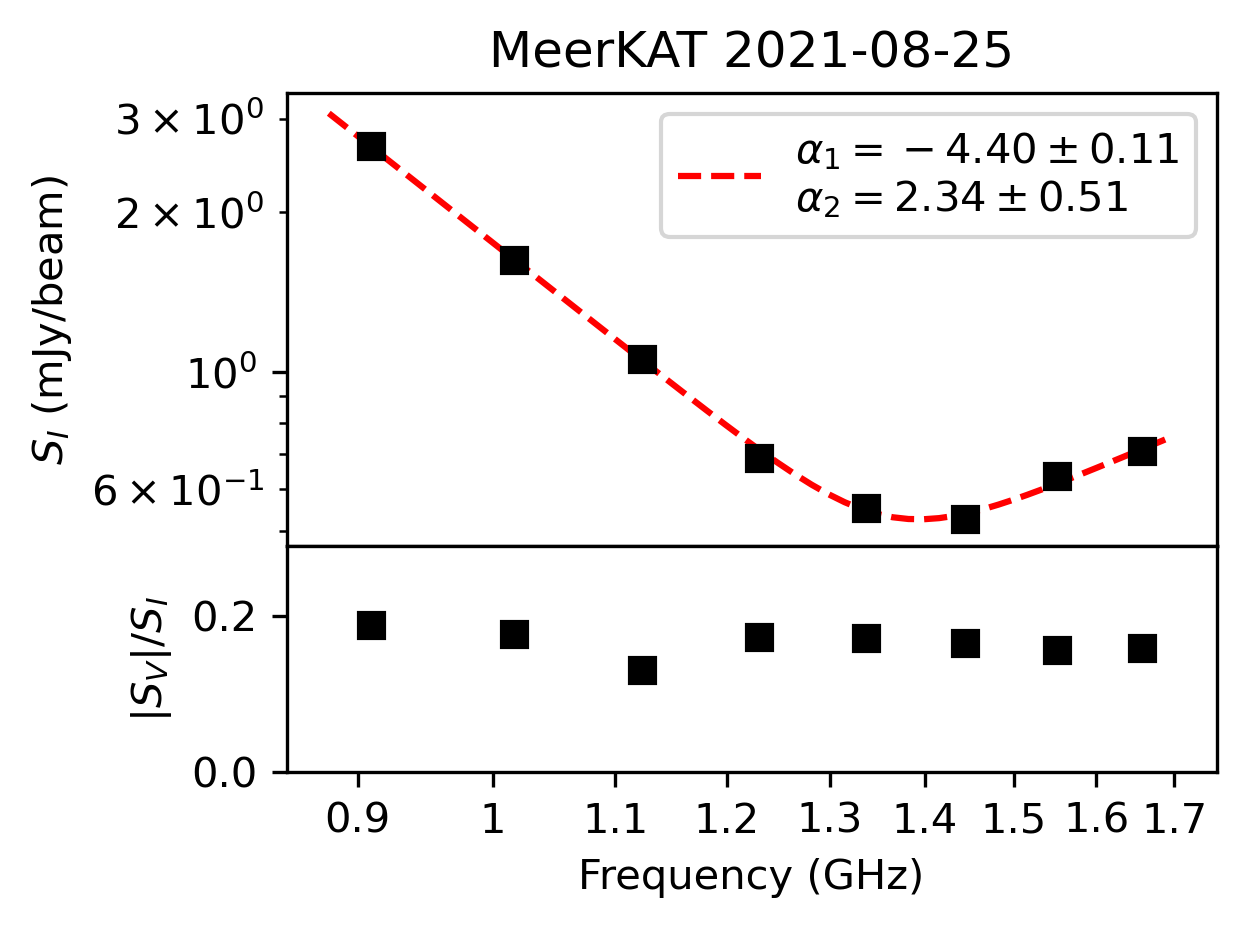}
	\includegraphics[width=0.46\textwidth]{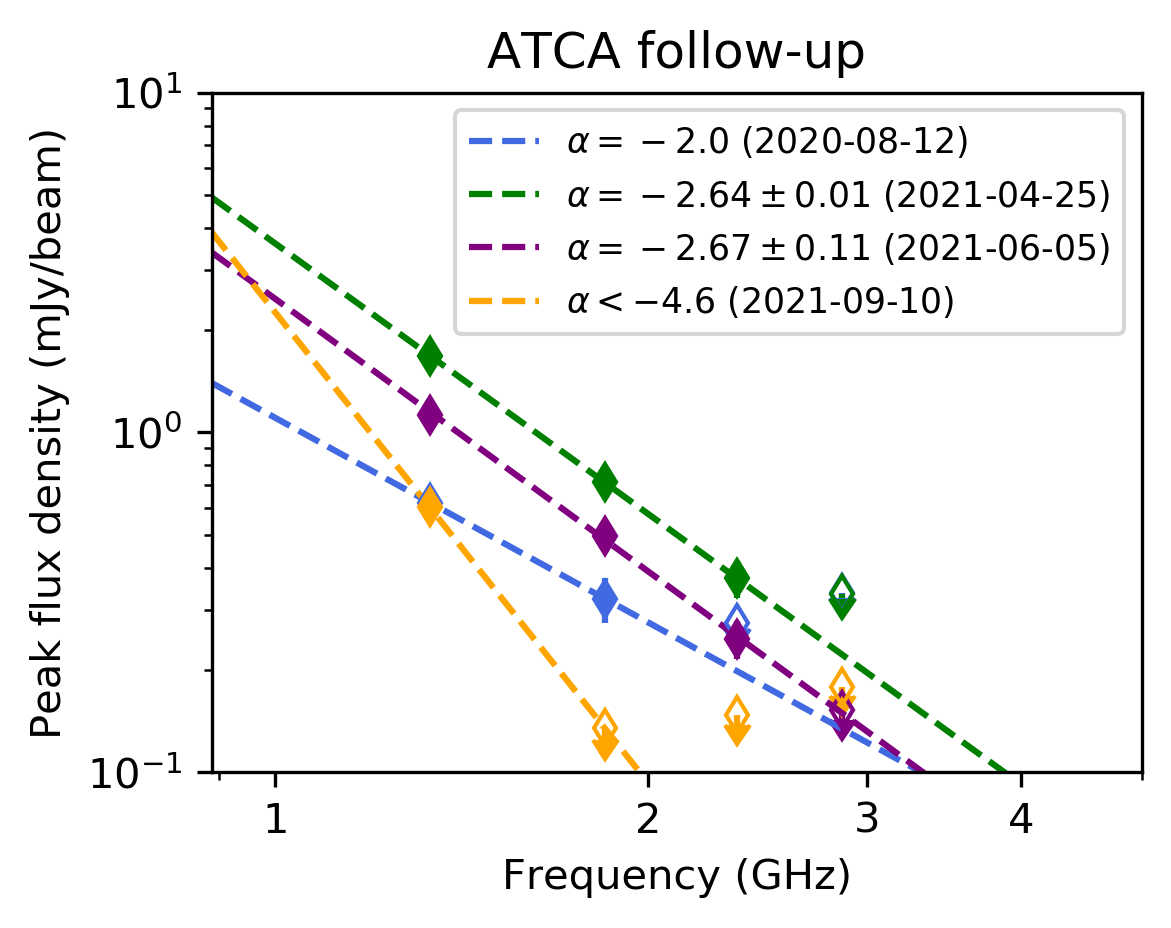}
    \caption{\textbf{Left:} The spectrum measured at the MeerKAT follow-up, with Stokes I (total intensity) in the upper panel and fractional circular-polarization ($|S_V|/S_I$) in the lower panel. The errorbars are smaller than the marker size. We split the 856\,MHz bandwidth into 8 sub-bands, and fitted with a smoothly broken power-law. \textbf{Right:} The spectra measured at different ATCA follow-up epochs, with errorbars smaller than the marker size. We equally split the 2\,GHz bandwidth into 4 sub-bands (centered at 1332\,MHz, 1844\,MHz, 2356\,MHz, and 2868\,MHz respectively). The open markers represent $5\sigma$ upper limits. We fit the spectra with a single power-law in each case, and the fitted spectral indices are shown in the legend. See Section~\ref{subsec:atca_archival_radio} and \ref{subsec:parkes_meerkat} for details. }
    \label{fig:spectral}
\end{figure*}

Prior to having identified the pulsations in \psr, we carried out follow-up observations of \source\ with the Australia Telescope Compact Array (ATCA) on 2020~August~12\footnote{The array was in a compact configuration EW352 on 2020~August~12, and only baselines including antenna~6 were included, corresponding to a baseline range of $4087-4439$\,km. },  
2021~April~25, 2021~June~4, and 2021~September~10 respectively, with a 2\,GHz bandwidth centered at 2.1\,GHz, 5.5\,GHz, and 9.0\,GHz (project code C3363, PI: Murphy; and project code C3431, PI: Pritchard). 
We reduced the data using \textsc{miriad}~\citep{1995ASPC...77..433S} with PKS B1934$-$638 and B0530$-$727 as the flux and phase calibrators respectively, and imaged the data using \textsc{casa}~\citep{2007ASPC..376..127M}. 
We detected the source with a flux density of 0.3$-$0.9\,mJy at 2.1\,GHz, but did not detect it at 5.5\,GHz or 9.0\,GHz in any epoch. 
We measured a circular polarization flux density of $0.076\pm0.013$\,mJy\,beam$^{-1}$ at 2.1\,GHz for the observation on 2021~June~4, implying a polarization fraction of $\sim$20\%. 
\added{The source is unresolved in all detected ATCA images, with angular resolution of 3--6$''$. }
We measured the spectral index by splitting the L-band (2.1\,GHz) into 4 sub-bands, and found it to be very steep, varying between $\alpha \sim -4$ to $\sim -2$. 
The spectra can be fitted using a single power-law, and their distributions are shown in Figure~\ref{fig:spectral}. 
The ATCA observations allowed us to measure a more accurate position compared to ASKAP ($\alpha=05^\mathrm{h}23^\mathrm{m}48.66^\mathrm{s}$, $\delta=-71^\circ25'52.58''$ in J2000 coordinates), with positional uncertainties of $\sim$0.15\,arcsec in each coordinate. 

We also detected \source\ in archival radio observations including the Sydney University Molonglo Sky Survey (SUMSS; \citealt{2003MNRAS.342.1117M}) and an ASKAP commissioning observation (SB8532; \citealt{2021MNRAS.506.3540P}). 
We obtained a higher-time resolution lightcurve from the latter (by imaging $9\times7$\,min scans separated with 1.2\,h), and found no significant variability within the 12\,h\,40\,m observation. 
\added{This area is also observed by the GaLactic and Extragalactic All-sky Murchison Widefield Array (GLEAM) survey \citep{Hurley-Walker2017}, and the non-detection yields a $3\sigma$ upper limit of $\sim$40\,mJy at 72--231\,MHz. }
All of these observations are listed in Table~\ref{tab:observations}.

\begin{deluxetable*}{cccccccccc}
\label{tab:observations}
\tablecaption{Summary of available observations for \source. Archival surveys giving best constraints are selected for optical/infrared. In the column of flux density, radio continuum observation refers to the total intensity (Stokes I) with units of mJy\,beam$^{-1}$, optical/infrared observation refers to the magnitude (Vega system), and X-ray observation refers to the unabsorbed flux (0.2--10\,keV). Flux densities are not available for Parkes pulsar observations.}
\tablewidth{0pt}
\tablehead{
\colhead{Obs. Date} & \colhead{Telescope} & \colhead{Duration} & \colhead{Band} & \colhead{Flux density} & \colhead{$|S_\textrm{V}|$} & \colhead{$|S_\textrm{V}|/S_\textrm{I}$} & \colhead{Survey/Project ID} & \colhead{Ref.} \\
\colhead{(UTC)} & \colhead{} & \colhead{} & \colhead{(MHz)} & \colhead{(mJy\,beam$^{-1}$)} & \colhead{(mJy\,beam$^{-1}$)} & &
}
\startdata
1997~Dec~22 & MOST & 11.5\,h & 843 & $8.6\pm0.9$ & \nodata & \nodata & SUMSS & 1 \\
2013~Nov~15 & MWA & $\sim$8\,m & 72--231 & $\lesssim40$ ($3\sigma$) & \nodata & \nodata & GLEAM & 2 \\
2019~Mar~13 & ASKAP & 12\,h\,40\,m & 1420 & $3.6\pm0.1$ & \nodata & \nodata & SB8178 \\
2019~Apr~20 & ASKAP & 12\,h\,40\,m & 888 & $4.3\pm0.1$ & \nodata & \nodata & EMU/SB8532 & 3 \\
2019~May~7 & ASKAP & 15\,m & 888 & $2.07\pm0.15$ & $<1.00$ & $<48\%$ & RACS & 4 \\ 
2019~Aug~28 & ASKAP & 12\,m & 888 & $2.44\pm0.31$ & $<1.24$ & $<51\%$ & VAST-P1 \\
2019~Dec~19 & ASKAP & 12\,m & 888 & $7.24\pm0.25$ & $1.38\pm0.22$ & 19\% & VAST-P1 \\
2020~Jan~10 & ASKAP & 12\,m & 888 & $4.73\pm0.25$ & $<1.04$ & $<22\%$ & VAST-P1 \\
2020~Jan~24 & ASKAP & 12\,m & 888 & $5.99\pm0.22$ & $1.48\pm0.21$ & 25\% & VAST-P1 \\
2020~Jan~25 & ASKAP & 12\,m & 888 & $5.13\pm0.21$ & $1.20\pm0.20$ & 23\% & VAST-P1 \\
2020~Apr~13 & Parkes & 30\,m & 704--4032 & \nodata & \nodata & \nodata & P1069 \\
2020~Jul~1 & Parkes & 60\,m & 704--4032 & \nodata & \nodata & \nodata & P1069 \\
2020~Aug~12 & ATCA & 2\,h & 2100 & $0.836\pm0.036$ & $<0.159$ & $<19\%$ & C3363, NAPA \\
2020~Aug~12 & ATCA & 2\,h & 5500 & $<0.122$ & \nodata & \nodata & C3363, NAPA \\
2020~Aug~12 & ATCA & 2\,h & 9000 & \nodata & \nodata & \nodata & C3363, NAPA \\
2020~Aug~29 & ASKAP & 12\,m & 888 & $3.89\pm0.24$ & $<1.06$ & $<27\%$ & VAST-P1 \\ 
2020~Nov~18 & Parkes & 135\,m & 704--4032 & \nodata & \nodata & \nodata & P1069 \\
2021~Apr~1 & ASKAP & 12\,m & 888 & $7.40\pm0.23$ & $1.47\pm0.19$ & 20\% & VAST-P2-low \\
2021~Apr~2 & ASKAP & 12\,m & 1367.5 & $1.47\pm0.26$ & $<1.31$ & $<89\%$ & VAST-P2-mid \\
2021~Apr~25 & ATCA & 80\,m & 2100 & $0.687\pm0.038$ & $<0.148$ & $<22\%$ & C3431 \\
2021~Apr~25 & ATCA & 60\,m & 5500 & $<0.128$ & \nodata & \nodata & C3431 \\
2021~Apr~25 & ATCA & 60\,m & 9000 & $<0.083$ & \nodata & \nodata & C3431 \\
2021~Jun~4 & ATCA & 4\,h & 2100 & $0.364\pm0.020$ & $0.076\pm0.013$ & 21\% & C3431 \\
2021~Jun~20 & ASKAP & 13\,m & 888 & $7.6\pm0.17$ & $1.5\pm0.18$ & 20\% & VAST-P2-low \\
2021~Aug~25 & MeerKAT & 2.5\,h & 1284 & $0.860\pm0.007$ & $0.137\pm0.005$ & 16\% & DDT-20210818-TM-01  \\
2021~Sep~10 & ATCA & 5\,h & 2100 & $0.176\pm0.018$ & \nodata & \nodata & C3431 \\
\tableline
1995 -- 1999 & LCST/GCC & \nodata & \textit{U} & $>20.7^\mathrm{m}$ & & & MCPS & 5 \\
 & & \nodata & \textit{B} & $>22.6^\mathrm{m}$ & & \\
 & & \nodata & \textit{V} & $>22.5^\mathrm{m}$ & & \\
 & & \nodata & \textit{I} & $>21.2^\mathrm{m}$ & & \\
2009 -- 2012 & VISTA & 7\,200\,s & \textit{Y} & $>21.1^\mathrm{m}$ & & & VMC & 6 \\
 & & 7\,200\,s & \textit{J} & $>20.9^\mathrm{m}$ & & \\
 & & 27\,000\,s & \textit{Ks} & $>20.4^\mathrm{m}$ & & \\
2021~May~29 & Swift/XRT & 2\,250\,s & 0.2--10\,keV & $<1.31\times10^{-13}$ & & & Target ID: 14338 \\
 & & & & (erg/s/cm$^2$) & 
\enddata
\tablecomments{
No Stokes V images/catalogues are available for SUMSS, ASKAP SB8178, or ASKAP SB8532; 
The non-detection upper limits of radio flux measurements are at $5\sigma$ confidence level; 
\added{For the three Stokes V non-detections in the VAST-P1, we combined these three epochs and measured a $3\sigma$ peak at the position of \source\ from the combined mean image, which yields a $|S_V|/S_I\approx15\%$, consistent with other detections; } 
The upper limits of other multi-wavelength observations are listed at the $3\sigma$ confidence level. 
Due to bad image quality for the ATCA observation at 9000\,MHz on 2020~August~12 (compact array configuration) and for the ATCA stokes V image on 2021~September~10 (strong artifacts), those images are discarded and not reported here. 
}
\tablerefs{
(1)~\citet{2003MNRAS.342.1117M};
(2)~\citet{Hurley-Walker2017};
(3)~\citet{2021MNRAS.506.3540P}; (4)~\citet{2020PASA...37...48M}; (5)~\citet{2004AJ....128.1606Z}; (6)~\citet{2011AA...527A.116C}}
\end{deluxetable*}

\subsection{Multiwavelength imaging observations}

Prior to initiating our pulsar searches we searched archival multiwavelength data including \textit{Gaia}~\citep{2018A&A...616A...1G}, the \textit{Wide-field Infrared Survey Explorer}~\citep[\textit{WISE};][]{2010AJ....140.1868W}, the \textit{ROSAT} All Sky Survey~\citep{2016A&A...588A.103B}, and the \textit{Fermi} Large Area Telescope source catalog~\citep{2015ApJS..218...23A}. 
No multiwavelength counterparts were found within a 5\,arcsec radius in \textit{Gaia} or \textit{WISE}, or within a 2\,arcmin radius in \textit{ROSAT} or \textit{Fermi}. 

The most sensitive infrared data available for this region is the VISTA Magellanic survey catalogue~\citep[VMC;][]{2011AA...527A.116C}. 
We found no sources within a $5\sigma$ position uncertainty, leading to 3$\sigma$ limiting magnitudes $Y>21.1$\,mag, $J>20.9$\,mag, and ${K_s}>20.4$\,mag in this region.
The position is also covered by the Magellanic Clouds Photometric Survey~\citep[MCPS;][]{2004AJ....128.1606Z}, and its non-detections of the source indicate $U>20.7$\,mag, $B>22.6$\,mag, $V>22.5$\,mag, and $I>21.2$\,mag. 

We conducted a Neil Gehrels \textit{Swift} Observatory Target of Opportunity (ToO) observation (target ID: 14338) on 2021 May 29 using the X-ray Telescope (XRT) in the photon counting mode (exposure time of 2250\,s). 
There is one count within a 15\,arcsec radius of \source, implying a count-rate upper limit of  $\sim0.0013$\,s$^{-1}$ (0.2--10\,keV). 
We therefore estimated a tentative upper limit to the unabsorbed flux of $\sim1.3\times10^{-13}$\,erg\,s$^{-1}$\,cm$^{-2}$ (0.2--10\,keV) based on an HI column density $\mathrm{N_H}=0.6\times10^{22}$\,cm$^{-2}$ (similar to the LMC magnetar SGR~0526$-$66; \citealt{2012ApJ...748..117P}) and a power-law photon index of $\Gamma=2$ using the HEASARC web-based PIMMS. 
Details of these multi-wavelength observations are summarised in Table~\ref{tab:observations}.

\section{Discussion} \label{sec:discussion}

\source\ was initially selected in the VAST continuum survey for follow-up due to its high variability and strong circular polarization ($f_p\sim20\%$), suggesting it is likely a pulsar or stellar object. 
The absence of any deep infrared/optical counterpart ($3\sigma$ limit $J>20.9$\,mag) strongly suggested it was a pulsar. 
Pulsar searches with MeerKAT and Parkes confirmed the pulsar nature of \source.
This pulsar, \psr, has a period of 322.5\,ms, DM of 157.5\,pc\,cm$^{-3}$, and wide pulse profile. 
Our ATCA observations show very steep spectral indices ranging from $\alpha\sim-4$ to $\sim-2$. 
The MeerKAT spectrum shows an unusual upturn at 1.36\,GHz from $\alpha=-4.4$ to $\alpha=2.34$. 
We will discuss properties of this pulsar in detail below. 

\subsection{Pulsar properties}
\label{subsec:pulsar_properties}


\begin{figure*}
    \centering
	\includegraphics[width=0.48\textwidth]{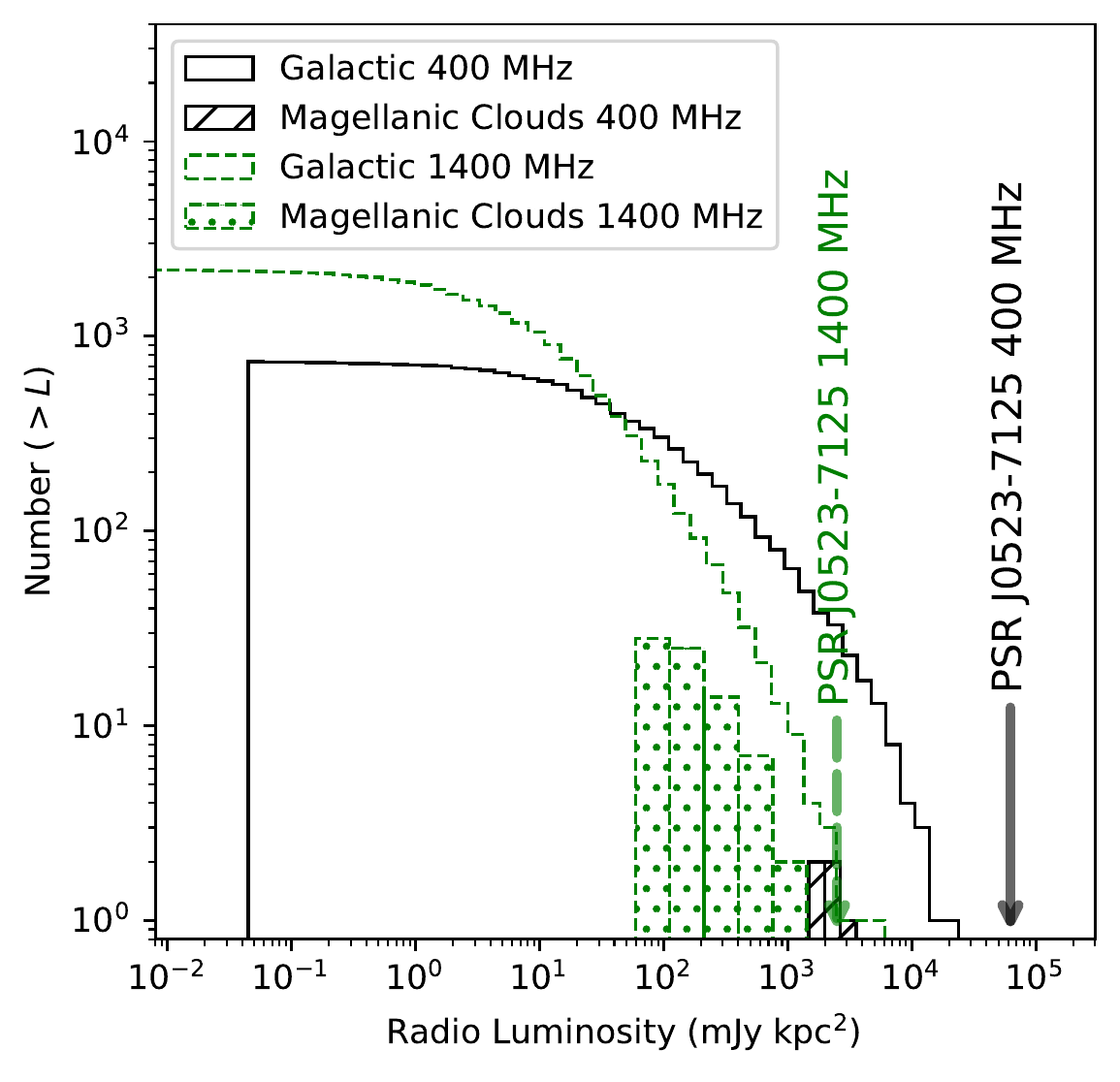}
	\includegraphics[width=0.5\textwidth]{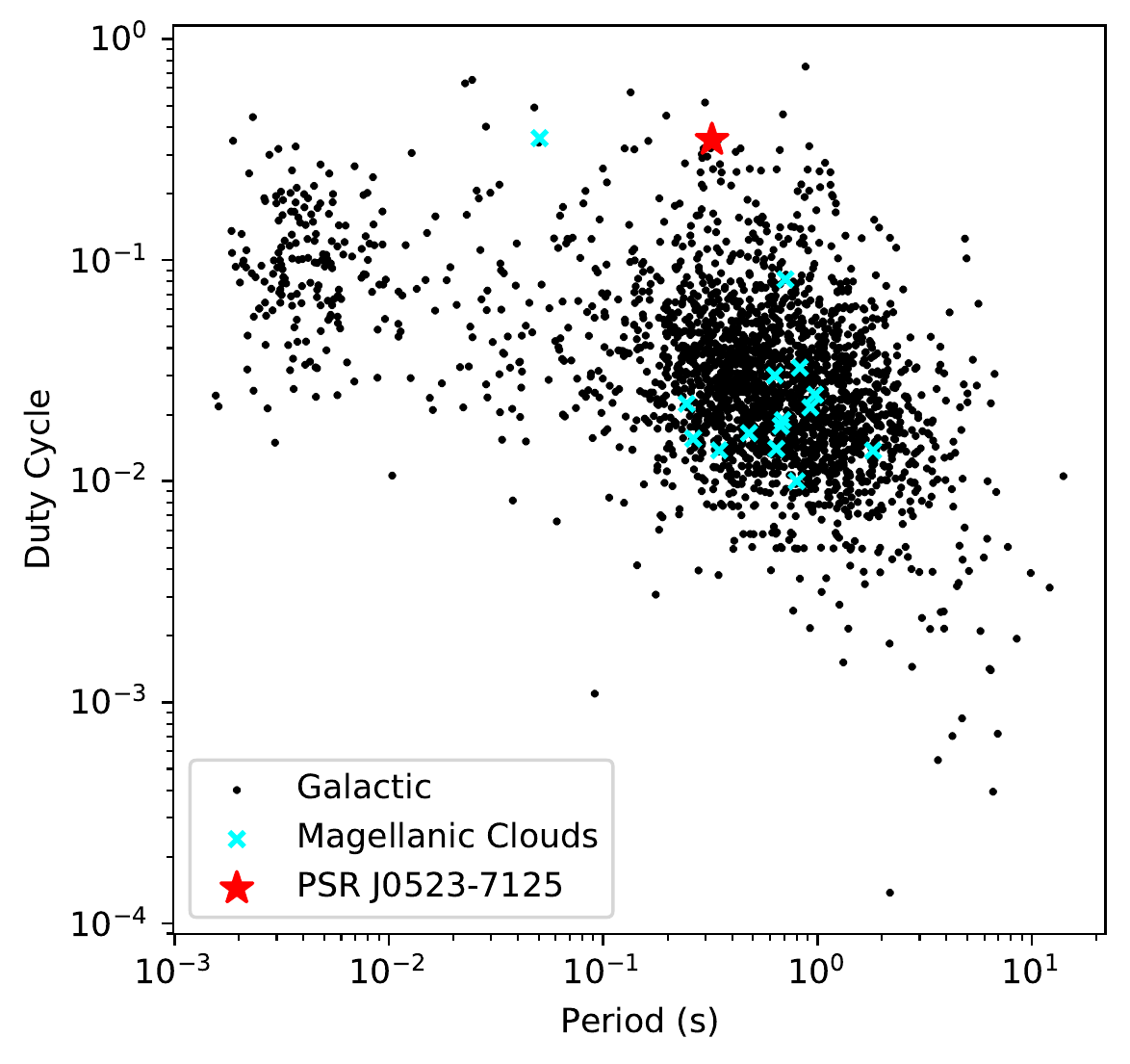}
    \caption{The left panel is the cumulative luminosity density function for known pulsars in the Galaxy \added{(unfilled)} and the Magellanic Clouds \added{(filled)}. The black color represents luminosity at 400\,MHz ($L_\mathrm{400}$), and green color represents luminosity at 1400\,MHz ($L_\mathrm{1400}$). 
    \added{The histogram filled with black stripes represents the luminosity distribution of Magellanic Clouds pulsars at 400\,MHz, and the histogram filled with green dots represents the luminosity distribution of Magellanic Clouds pulsars at 1400\,MHz. }
    The right panel is period versus duty cycle distribution of known pulsars, with Galactic pulsars marked as black and Magellanic Clouds pulsars marked as cyan. The red star represents our pulsar, \psr, which shows an offset from the majority of the pulsar population (as a slow pulsar with large duty cycle). 
    Data is from the ATNF pulsar catalogue \citep[][catalog version 1.66]{2005AJ....129.1993M}. }
    \label{fig:pulsar_distribution}
\end{figure*}

As shown in Figure~\ref{fig:vast_cutout}, we know the source is located in the direction of the LMC. 
The maximum DM contribution from the Milky Way along this line-of-sight is  $\sim50\textrm{--}60$\,pc\,cm$^{-3}$ (NE2001 and YMW16; \citealt{2002astro.ph..7156C}; \citealt{2017ApJ...835...29Y}), and there is no known Galactic HII region along this line-of-sight. 
With a DM of 157.5\,pc\,cm$^{-3}$, \psr\ almost certainly resides in the LMC. 
We also note this DM is consistent with the DMs of other pulsars found in the LMC ($\sim50\textrm{--}270$\,pc\,cm$^{-3}$; e.g., \citealt{1991MNRAS.249..654M, 2006ApJ...649..235M, Ridley2013, Johnston2022}). 

We can calculate the monochromatic radio luminosity density of \psr\ at 400\,MHz and 1400\,MHz respectively ($L_\nu=S_\nu d^2$, where $S_\nu$ is the flux density and $d$ is the pulsar distance). 
Our ATCA, MeerKAT, and ASKAP mid-band observations covered 1400\,MHz, and the measured average flux density is  $S_\mathrm{1400}\approx1$\,mJy. 
The flux density at 400\,MHz is therefore $S_\mathrm{400}\approx25$\,mJy assuming a median spectral index $\alpha\sim-2.6$ (Figure~\ref{fig:spectral}). 
The flux density at 400\,MHz would be ten times higher for the steepest spectral index that we measure ($\alpha\sim-4.4$).
The pulsar distance $d$ is assumed to be the LMC distance of 50\,kpc \citep{Pietrzynski2013}, which is reasonable since the uncertainty in luminosity should be dominated by the uncertainty in the flux density. 
We can then estimate $L_\mathrm{400}\approx6.3\times10^4$\,mJy\,kpc$^2$ and  $L_\mathrm{1400}\approx2.5\times10^3$\,mJy\,kpc$^2$, which is one of the most luminous pulsar to date (see Figure~\ref{fig:pulsar_distribution}). 
Only one known pulsar with a well-constrained distance, B1641$-$45~\citep{Komesaroff1973, Frail1991}, has a comparable luminosity to \psr\ ($\sim7.6\times10^3$\,mJy\,kpc$^2$ at 400\,MHz and $\sim6.1\times10^3$\,mJy\,kpc$^2$ at 1400\,MHz). 
We exclude other high-luminosity pulsars, like PSR~J0134$-$2937, from consideration due to their significant distance uncertainties. 

\citet{Ridley2013} calculated the luminosity function of the LMC pulsars, and noticed a discrepancy between the Galactic and LMC luminosity distributions. 
They attributed the discrepancy to a bias due to the small sample size of the LMC pulsars, and concluded that the luminosity function for the LMC pulsars is consistent with its counterpart in the Galactic disc. 
They also suggested that the maximum 1400\,MHz radio luminosity for LMC pulsars is approximately $1000$\,mJy\,kpc$^2$ based on their sample. 
Our pulsar is about a factor of $\sim$2 more luminous than this limit. 
With the discovery of \psr, we can revisit the luminosity function of the LMC pulsars by fitting the high-luminosity tail of the distribution as a power law, i.e., $\log N = F\log L + G$, where $F$ is the slope of the distribution \citep{Lorimer2006}. 
Based on the ATNF pulsar catalog~\citep{2005AJ....129.1993M}, we estimate $F\approx-1.30\pm0.03$ for Galactic pulsars with luminosities above 30\,mJy\,kpc$^2$ (consistent with other work, e.g., \citealt{Lorimer2004, Ridley2013}), and $F\approx-1.6\pm0.2$ for LMC pulsars with luminosities above 125\,mJy\,kpc$^2$. 
Although the calculation is still limited by small-number statistics for the LMC pulsars and the distance uncertainties for Galactic pulsars, our result shows that the LMC pulsars are consistent with the luminosity function of their Galactic counterparts (as suggested by \citealt{2006ApJ...649..235M, Ridley2013}). 

\psr\ has a wide pulse profile with $W_\mathrm{50}\approx100$\,ms at $\sim$1\,GHz.
Figure~\ref{fig:pulsar_distribution} shows the duty cycle distribution of known pulsars, and \psr\ generally has a larger duty cycle than others (either Galactic or extragalactic pulsars). 
The observed pulse width $W$ can be decomposed as $W=\sqrt{\tau_\mathrm{int}^2+\tau_\mathrm{sc}^2+\tau_\mathrm{DM}^2}$, where $\tau_\mathrm{int}$ is the intrinsic pulse width, $\tau_\mathrm{sc}$ is the pulse broadening due to multipath scattering, and $\tau_\mathrm{DM}$ is the pulse broadening due to dispersion \citep{2012hpa..book.....L}. 
Both $\tau_\mathrm{sc}$ and $\tau_\mathrm{DM}$ are related to free electrons in the interstellar medium \citep[i.e., the DM;][]{2004ApJ...605..759B}.
Following the YMW16 model (which includes the distribution of free electrons in the Galaxy and the Magellanic Clouds), the predicted scattering delay for Magellanic Clouds pulsars along this line-of-sight is $\tau_\mathrm{sc}\ll1$\,ms at 1100\,MHz. 
This suggests that the wide pulse profile is largely intrinsic. 
Noted in Section~\ref{subsec:parkes_meerkat}, the pulse width may be further underestimated due to the difficulty of identifying off pulse baseline. 
If this is the case, \psr\ could potentially have a nearly 100\% duty cycle, i.e., a possible nearly-aligned rotator (the magnetic and rotation axis of the star are aligned; e.g., \citealt{2010MNRAS.402.1317Y}). 
This assumption could also explain why the pulsar is so bright in continuum images, but hard to detect in a pulsar search. 
The pulsar calibration is ongoing and we will discuss it further in a later paper. 

Our MeerKAT pulsar data measured a $\textrm{RM}=+456$\,rad\,m$^{-2}$, and the Parkes data confirmed this result. 
However, we cannot find any significant linear polarization in MeerKAT continuum data in a RM range from $-1200$ to $1200$\,rad\,m$^{-2}$. 
This discrepancy can be attributed to potential swing of the polarization position angle in the pulse phase \added{(e.g., following the characteristic S-shaped curve of the `rotating vector model'; \citealt{Radhakrishnan1969})}, making the effective linear polarization fraction very diluted when integrated in the continuum image. 
Further precise polarimetry can confirm this. 
We noticed this $|\mathrm{RM}|$ is about twice the most extreme $|\mathrm{RM}|$ of known LMC pulsars, PSR~J0540$-$6919 with $\textrm{RM}=-246$\,rad\,m$^{-2}$ \citep{Johnston2022}, which is located in a supernova remnant. 
We checked the UM/CTIO Magellanic Cloud emission-line survey (MCELS; \citealt{Smith1998}), and found there is some faint, diffuse emission around our pulsar in the H$\alpha$ map (with an angular scale $\sim$13\,arcmin). 
We got a H$\alpha$ intensity $I_\mathrm{H\alpha}\approx1.3\times10^{-16}$\,erg\,cm$^{-2}$\,s$^{-1}$\,arcsec$^{-2}$ from the MCELS map after extinction correction (following the method in \citealt{Gaensler2005}), and thus the emission measure (EM) is $\sim$70\,pc\,cm$^{-6}$. 
The foreground contribution to the DM in this line-of-sight is about 60\,pc\,cm$^{-3}$ \citep{2017ApJ...835...29Y}, and to the RM is about $+31$\,rad\,m$^{-2}$ \citep{Mao2012}. 
After subtracted the Galactic foreground contribution, the estimated parallel magnetic field $B_\parallel=1.23\,\mathrm{RM_{LMC}}/\mathrm{DM_{LMC}}\approx5\,\mu$G, and the estimated electron column density $n_e=\mathrm{EM}/\mathrm{DM_{LMC}}\approx0.7$\,cm$^{-3}$. 
The occupation length, $fL$, of ionized gas is $\mathrm{DM_{LMC}}^2/\mathrm{EM}\approx150$\,pc, where $L$ is the projected length and $f$ is the filling factor (the fraction of the line-of-sight for free electron density). 
For reasonable values of $f$, $L/d$ is of the order of the angular scale of the H$\alpha$ emission (where $d=50$\,kpc for the LMC). 
If we assume the angular size is $13$\,arcmin (as what we observed from the MCELS image), the inferred filling factor $f\approx0.8$ and the length $L\approx180$\,pc. 
Those values are reasonable for an evolved HII region.
If the pulsar turns out to be young in future timing analysis, it is likely embedded in this diffuse gas region. 

In our continuum observations, \source\ shows strong variability with a modulation index $\sim$43\% at 888\,MHz over timescales of $\sim$days (see the lightcurve in Figure~\ref{fig:lightcurve}). 
There is no significant continuum variability within the 2.5\,h MeerKAT observation or the 12\,h ASKAP EMU observation. 
A structure-function analysis shows that the variability timescale is $\sim$17\,d, though this is poorly constrained due to the absence of samples between 1\,d and 14\,d. 
This strong variability could be intrinsic (e.g., pulsar nulling; \citealt{1970Natur.228...42B}) or related to propagation effects (e.g., diffractive scintillation; \citealt{1990ARA&A..28..561R}). 
The diffractive scintillation bandwidth follows $\Delta f_\mathrm{DISS} \sim C_1/2\pi\tau_\mathrm{sc}$\,MHz \citep{2002astro.ph..7156C}, where $C_1=1.16$ assuming a Kolmogorov spectrum and $\tau_\mathrm{sc}$ is the scattering time. 
The predicted scattering delay for the LMC pulsars in this line-of-sight is $\tau_\mathrm{sc}\sim10^{-2}$\,ms based on the YMW16 model, and thus $\Delta f_\mathrm{DISS}\sim10^{-2}$\,MHz. 
The scintillation strength, $u=\sqrt{f/\Delta f_\mathrm{DISS}}$ where $f$ is the observing frequency, is therefore $u\gg1$, implying a strong scattering regime (i.e., diffractive scintillation and/or refractive scintillation).
Diffractive scintillation in this case is an unlikely explanation, as the calculated $\Delta f_\mathrm{DISS}$ is about $10^{-2}$\,MHz, far lower than the observing frequency channel width $\sim1$\,MHz. 
For refractive scintillation, the calculated modulation index $m=\sqrt{u^{5/3}}\sim16\%$, and the timescale $\sim10^2$\,day. 
\citet{1998MNRAS.294..307W} estimated the effects of interstellar scintillation (arise from the Milky Way) on extragalactic source using the TC93 model~\citep{Taylor1993}, and we find a different modulation index of $m\sim42\%$ and timescale of $\sim2$\,day based on this model.  
These estimations are not too far from our observed results, we therefore think the variability can be explained by propagation effects. 
However, we cannot rule out the possibility of contribution from intrinsic variation. 
As we mentioned before, the large duty cycle of the pulsar suggests that the pulsar could be an aligned rotator, which are known to show larger levels of nulling~\citep{Cordes2008}. 
Aligned rotating pulsars also show mode changing behaviour (e.g. PSR J1107-5907, \citealt{Hobbs2016}).

The source shows very steep spectral indices (varying from $\alpha\sim-4$ to $\alpha\sim-2$) even compared to the pulsar population (average $\alpha$ of $-1.4$; \citealt{2013MNRAS.431.1352B}, and $\alpha<-2.5$ for the fastest rotating millisecond pulsar; e.g., \citealt{2016ApJ...829..119F}). 
The high quality MeerKAT continuum data show an unusual radio spectral shape, with an upturn at 1.36\,GHz from $\alpha=-4.4$ to $\alpha=2.34$. 
The upturn at $\sim1$\,GHz is hard to explain as most pulsars can be described using a simple power law spectrum.
Some pulsar spectra are known to show a turnover at low frequencies due to synchrotron self-absorption or thermal free-free absorption, however this is the opposite of what we see (i.e., transitioning from a flat to a steeper spectrum; see examples in \citealt{2018MNRAS.473.4436J}). 
We also note that some pulsars can have an upturn spectrum at higher frequency, but this usually occurs at millimeter wavelengths ($\gtrsim10$\,GHz; \citealt{Kramer1996}). 
Therefore, we instead consider external effects. 
\citet{2017MNRAS.469.5023T} found that scintillation can cause kinks, bumps and wiggles in the broad-band radio spectrum of a quasar. 
If \psr\ is scintillating (as we suggested above), we might be able to see unusual spectra structure like these. 

Further observations should be able to conclusively
determine a timing solution, providing an age and spin-down luminosity of the source. 
We will discuss these properties in a later paper.

\subsection{Identifying pulsars in continuum surveys}
\label{sec:continuum_survey}


The Magellanic Clouds have been targeted several times with pulsar surveys using the Parkes telescope~\citep[e.g.,][]{McCulloch1983, 1991MNRAS.249..654M, Crawford2001, 2006ApJ...649..235M, Ridley2013}, identifying 31 extragalactic pulsars. 
\psr\ is brighter than all of these, but was not identified in these surveys. 
Its wide pulse profile and steep spectrum could be responsible for its non-detection since they would reduce the signal-to-noise ratio. 
This is especially true in surveys with the Parkes Multi-Beam receiver at 1400\,MHz, where many recent searches have been conducted. 
As described in Section~\ref{subsec:parkes_meerkat}, we cannot easily detect the pulsar using traditional methods from targeted Parkes data. 
However, we can clearly identify it in Parkes using only low-frequency data from the UWL receiver, consistent with the spectral properties determined from continuum images. 

There are many continuum-based pulsar searches using a steep spectral index as the selection metric \citep[e.g.,][]{1985PASA....6..174D, 2000ApJ...535..975C, 2018ApJ...864...16M, 2019ApJ...876...20H}, but only a limited number of new pulsars have been found \citep[e.g.,][]{Bhakta2017, Frail2018}. 
The steep-spectrum selection normally requires two observations at different frequencies to measure the spectral index. 
Moreover, there are also other types of sources that can have very steep spectra, such as high-redshift radio galaxies \citep{O'Dea1998}.
Circularly-polarized sources without deep infrared/optical counterparts are highly likely to be pulsars. 
We also note that such sources may belong to an as-yet unknown class or classes. 
For instance, the Galactic Center Radio Transient (GCRT; \citealt{2002AJ....123.1497H,2005Natur.434...50H}), 
two steep-spectrum, polarized sources near the Galactic bulge (C1748$-$2827 and C1709$-$3918; \citealt{2021MNRAS.507.3888H}), and a polarized transient recently discovered in the VAST survey near the Galactic Centre (\sourceandy; \citealt{2021ApJ...920...45W}) are circularly-polarized sources without deep infrared/optical counterparts, whose natures are still unknown. 

\begin{figure}
    \centering
	\includegraphics[width=0.8\columnwidth]{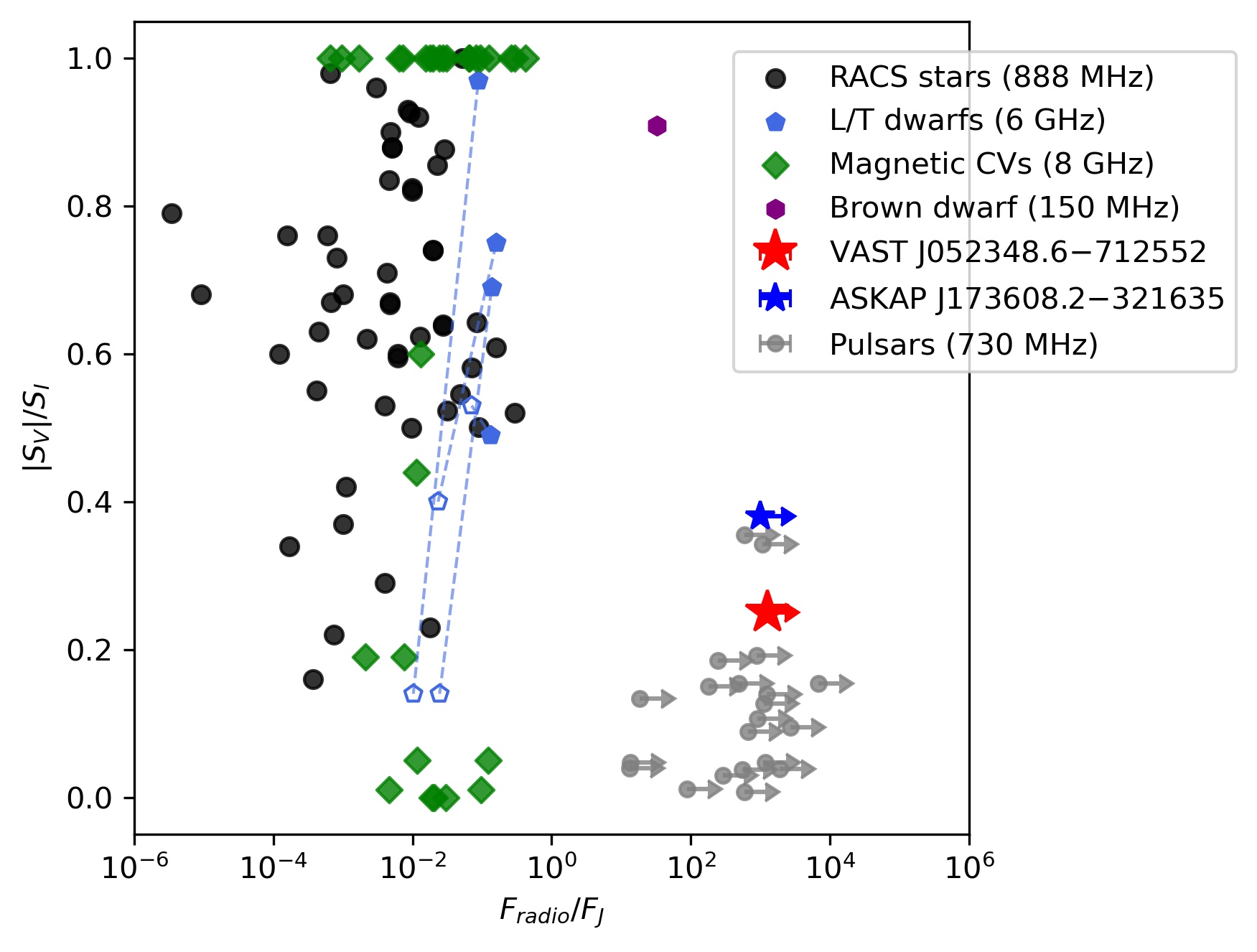}
    \caption{Fractional circular polarization versus radio-to-near-infrared flux ratio for stellar objects and pulsars. The black circles represent radio stars measured in RACS at 888\,MHz~\citep{2021MNRAS.502.5438P}, the blue pentagons represent quiescence (open symbols) and peak (filled symbols) auroral emissions from L/T dwarfs at 6\,GHz~\citep{2016ApJ...818...24K}, the green diamonds represent magnetic cataclysmic variables at 8\,GHz~\citep{2020AdSpR..66.1226B}, the purple hexagon represents the recently discovered brown dwarf BDR~J1750$+$3809 at 150\,MHz~\citep{2020ApJ...903L..33V}, the gray circles represent a group of pulsars measured at 730\,MHz~\citep{2015MNRAS.449.3223D}, the blue star represents another polarized transient \sourceandy\ detected in VAST-P1~\citep{2021ApJ...920...45W}, and the red star represents \source. The near-infrared data were taken from various surveys with the Visible and Infrared Survey Telescope for Astronomy~\citep[VISTA;][]{2006SPIE.6269E..0XD,2006Msngr.126...41E} and 2-micron All Sky Survey~\citep[2MASS;][]{2006AJ....131.1163S}. We note that \source\ is consistent with the pulsar region. }
    \label{fig:star_limit}
\end{figure}

There are an increasing number of large-scale radio continuum surveys, ranging from low-frequency surveys such as the Low-Frequency Array (LOFAR) Two-metre Sky Survey (LoTSS; \citealt{Shimwell2017}), the Giant Metrewave Radio Telescope (GMRT) 150 MHz All-Sky Survey (TGSS; \citealt{Intema2017}), and the GaLactic and Extragalactic All-sky Murchison Widefield Array survey (GLEAM; \citealt{Hurley-Walker2017}), to gigahertz-frequency surveys such as the ASKAP Rapid Commissioning Survey (RACS; \citealt{2020PASA...37...48M}), the Evolutionary Map of the Universe survey with ASKAP (EMU; \citealt{Norris2021}), the Polarisation Sky Survey of the Universe's Magnetism (POSSUM; using the same sky coverage as EMU with full polarization; \citealt{Gaensler2010}), and the Karl G. Jansky Very Large Array (VLA) Sky Survey (VLASS; \citealt{Lacy2020}). 
With improved instruments in the Square Kilometre Array era, instantaneous large fields-of-view and great sensitivity will be even more common, leading to the detection of large numbers of radio sources across the sky. 
Some or most of the surveys include circular polarization measurements. 
Apart from radio wavelengths, there are significant improvements in multi-wavelengths surveys (with better sensitivity and large sky coverage), e.g., the VISTA Variables in the Via Lactea (VVV; \citealt{Minniti2010}), the VISTA Hemisphere Survey (VHS; \citealt{McMahon2013}), the Dark Energy Survey (DES; \citealt{Abbott2018des}), and the \textit{Gaia} mission \citep{GaiaCollaboration2016}. 
These deep multi-wavelength surveys can greatly help with the classification of a circular-polarized object (e.g., to check if there is any stellar counterpart). 
As shown in Figure~\ref{fig:star_limit}, which listed known circular polarized objects including various stellar objects, the radio-to-near-infrared flux ratio (plus the fractional circular polarization) can be used as a good diagnostic tool to select strong pulsar candidates. 

One limitation of circular-polarization selection is that not all pulsars have a high level of circular polarization (the median fraction $\sim$10\%; \citealt{Han1998, Johnston2018}). 
As estimated by \citet{2019ApJ...884...96K}, even deep ASKAP continuum searches (e.g., EMU, with sensitivity $\sim$50\,$\mu$Jy) for circularly polarized source are unlikely to find large numbers of new pulsars. 
However, they may be critical to identify extreme pulsars that are missed in traditional pulsar surveys. 
For example, the high electron density in the Galactic Centre direction (which has much higher stellar densities) causes strong scattering and makes pulsars relatively difficult to discover through traditional periodicity searches. 

Extragalactic pulsar searches in continuum images are also possible. 
To date the Magellanic Clouds are still the only place that extragalactic pulsars have been detected, even after several attempts searching for pulsars in other galaxies including M31, M33, and nearby dwarf galaxies \citep[e.g.,][]{McLaughlin2003, Bhat2011, Rubio-Herrera2013, Mikhailov2016, vanLeeuwen2020}. 
We can estimate the number of detectable pulsars in M31 (our neighbor galaxy) through continuum surveys, by putting all known Galactic pulsars at the distance of M31 ($785\pm25$\,kpc; \citealt{McConnachie2005}). 
We assume that the number of pulsars in M31 and our Galaxy is similar, though some work suggests there may be a smaller radio pulsar population in M31 \citep{vanLeeuwen2020}. 
We consider using the Next Generation Very Large Array (ngVLA; \citealt{Murphy2018}), which is about 10 times sensitive than the VLA and can achieve a resolution at $\sim$mas level. 
The ngVLA is located in the Northern hemisphere and therefore also ideal for observing M31 (and M33). 
The rms sensitivity of the ngVLA is expected to be $\sim$0.1\,$\mu$Jy\,beam$^{-1}$ for a 10\,h observation at $\sim1\textrm{--}3$\,GHz. 
Assuming this sensitivity, more than $\sim$50 simulated M31 pulsar could be detected at the $>5\sigma$ level in a survey of M31, and $\sim$5 could be detected in circular polarization. 
Although the estimated number is still very low, it is a promising possibility to detect the most luminous pulsars in M31 (e.g., those similar to \psr). 
In such an observation we would also be able to measure their flux densities and spectral properties, providing valuable information for a follow-up targeted periodicity searches. 

\section{Conclusion}

We discovered a highly-variable, circularly-polarized object, \source, in a variability analysis of two fields containing the Magellanic Clouds observed as part of the VAST-P1 survey. 
With $\sim$20\% fractional circular polarization and no optical/infrared counterpart, \source\ was considered to be a strong pulsar candidate. 
Subsequent MeerKAT observations discovered a pulsar, \psr, associated with the continuum source, which was further confirmed with observations using the Parkes telescope. 
The pulsar has a period of 322.5\,ms and a DM of 157.5\,pc\,cm$^{-3}$, consistent with a LMC origin. 
The extragalactic distance makes \psr\ among the most luminous known pulsars for steady-state emission; in particular it is brighter than all known pulsars in the Magellanic Clouds at both 400\,MHz and 1400\,MHz. 
Despite its high luminosity, \psr\ remained undetected in several LMC pulsar surveys, which we suspect is largely due to its wide pulse profile and/or steep-spectral shape. 
The wide pulse profile also suggests that \psr\ could be an aligned rotator. 
\added{The strong variability is likely due to scintillation effects, though we cannot rule out the possibility of any intrinsic variation. }
We measured a large $\mathrm{RM}=+456$\,rad\,m$^{-2}$ for this pulsar, which is about twice the most extreme $|\mathrm{RM}|$ of LMC pulsars. 
A preliminary analysis show that \psr\ is likely embedded in an evolved HII region, which would be further strengthened if it turns out to be a young pulsar in ongoing timing analysis. 

Our discovery highlights the possibility of identifying pulsars (especially non-standard pulsars) from continuum images, particularly when circular polarization is combined with (largely archival) multiwavelength data. 
Improved next generation radio telescopes and increasing number of large-scale multi-wavelength surveys will bring large amounts of data with great sensitivity and resolution, giving us an unprecedented opportunity to identify more pulsar (even for extragalactic pulsars farther than the Magellanic Clouds) via continuum images. 

\begin{acknowledgments}
We thank the anonymous reviewer for their useful comments. 
We thank Elaine Sadler, Ron Ekers, Elizabeth Mahony, Shami Chatterjee, Mark Walker, Keith Bannister, Natasha Hurley-Walker and Aris Karastergiou for useful discussions. 
We thank the MeerKAT and Swift directors for approving our DDT observations, and S.\ Buchner for assistance in scheduling and conducting MeerKAT observations.
YW is supported by the China Scholarship Council. 
TM acknowledges the support of the Australian Research Council (ARC) through grant DP190100561. 
DK and AO are supported by NSF grant AST-1816492.
VG, NR, and MR acknowledge support from the H2020 ERC Consolidator Grant ``MAGNESIA'' under grant agreement Nr.\ 817661 (PI: Rea), and grants SGR2017-1383 and PGC2018-095512-B-I00.
RMS acknowledges support through ARC Future Fellowship FT 190100155. 
AR gratefully acknowledges financial support by the research grant ``iPeska'' (PI: Andrea Possenti) funded under the INAF national call Prin-SKA/CTA approved with the Presidential Decree 70/2016 and continuing valuable support from the Max-Planck Society. AR also acknowledges support from the Ministero degli Affari Esteri e della Cooperazione Internazionale - Direzione Generale per la Promozione del Sistema Paese - Progetto di Grande Rilevanza ZA18GR02.
GRS is supported by NSERC Discovery Grants RGPIN-2016-06569 and RGPIN-2021-0400.
This research was supported by the Sydney Informatics Hub (SIH), a core research facility at the University of Sydney.
Parts of this research were conducted by the Australian Research Council Centre of Excellence for Gravitational Wave Discovery (OzGrav), project number CE170100004. 
This research has made use of the VizieR catalogue access tool, CDS, Strasbourg, France \citep{2000A&AS..143...23O}. 
The Australia Telescope Compact Array and the Parkes radio telescope are part of the Australia Telescope National Facility (grid.421683.a) which is funded by the Australian Government for operation as a National Facility managed by CSIRO.
The Australian SKA Pathfinder is part of the Australia Telescope National Facility which is managed by CSIRO. Operation of ASKAP is funded by the Australian Government with support from the National Collaborative Research Infrastructure Strategy. ASKAP uses the resources of the Pawsey Supercomputing Centre. Establishment of ASKAP, the Murchison Radio-astronomy Observatory and the Pawsey Supercomputing Centre are initiatives of the Australian Government, with support from the Government of Western Australia and the Science and Industry Endowment Fund. We acknowledge the Wajarri Yamatji people as the traditional owners of the Observatory site.
The MeerKAT telescope is operated by the South African Radio Astronomy Observatory, which is a facility of the National Research Foundation, an agency of the Department of Science and Innovation. 
\end{acknowledgments}

\vspace{5mm}
\facilities{ASKAP, ATCA, MeerKAT, Parkes, Swift. }

\software{
APLpy \citep{Robitaille2012}, 
Astropy \citep{2013A&A...558A..33A,2018AJ....156..123A}, 
CASA \citep{2007ASPC..376..127M}, 
CubiCal~\citep{2018MNRAS.478.2399K},
matplotlib~\citep{Hunter2007}, 
MIRIAD~\citep{1995ASPC...77..433S}, 
numpy~\citep{harris2020array}, 
oxkat~\citep{2020ascl.soft09003H}, 
PRESTO \citep{Ransom2011}, 
PSRCHIVE \citep{2004PASA...21..302H}, 
pulsar\_miner\footnote{\url{https://github.com/alex88ridolfi/PULSAR\_MINER}}, 
PyPulse \citep{2017ascl.soft06011L}, 
RM-Tools \citep{Purcell2020}, 
tricolour\footnote{\url{https://github.com/ska-sa/tricolour}}, 
wsclean~\citep{2014MNRAS.444..606O}. 
}

\bibliography{sample631}{}

\begin{thebibliography}{}
\expandafter\ifx\csname natexlab\endcsname\relax\def\natexlab#1{#1}\fi
\providecommand{\url}[1]{\href{#1}{#1}}
\providecommand{\dodoi}[1]{doi:~\href{http://doi.org/#1}{\nolinkurl{#1}}}
\providecommand{\doeprint}[1]{\href{http://ascl.net/#1}{\nolinkurl{http://ascl.net/#1}}}
\providecommand{\doarXiv}[1]{\href{https://arxiv.org/abs/#1}{\nolinkurl{https://arxiv.org/abs/#1}}}

\bibitem[{{Abbott} {et~al.}(2018){Abbott}, {Abdalla}, {Allam}, {Amara},
  {Annis}, {Asorey}, {Avila}, {Ballester}, {Banerji}, {Barkhouse}, \&
  et~al.}]{Abbott2018des}
{Abbott}, T.~M.~C., {Abdalla}, F.~B., {Allam}, S., {et~al.} 2018, \apjs, 239,
  18, \dodoi{10.3847/1538-4365/aae9f0}

\bibitem[{{Acero} {et~al.}(2015){Acero}, {Ackermann}, {Ajello}, {Albert},
  {Atwood}, {Axelsson}, {Baldini}, {Ballet}, {Barbiellini}, {Bastieri},
  {Belfiore}, {Bellazzini}, {Bissaldi}, {Blandford}, {Bloom}, {Bogart},
  {Bonino}, {Bottacini}, {Bregeon}, {Britto}, {Bruel}, {Buehler}, {Burnett},
  {Buson}, {Caliandro}, {Cameron}, {Caputo}, {Caragiulo}, {Caraveo},
  {Casandjian}, {Cavazzuti}, {Charles}, {Chaves}, {Chekhtman}, {Cheung},
  {Chiang}, {Chiaro}, {Ciprini}, {Claus}, {Cohen-Tanugi}, {Cominsky}, {Conrad},
  {Cutini}, {D'Ammando}, {de Angelis}, {DeKlotz}, {de Palma}, {Desiante},
  {Digel}, {Di Venere}, {Drell}, {Dubois}, {Dumora}, {Favuzzi}, {Fegan},
  {Ferrara}, {Finke}, {Franckowiak}, {Fukazawa}, {Funk}, {Fusco}, {Gargano},
  {Gasparrini}, {Giebels}, {Giglietto}, {Giommi}, {Giordano}, {Giroletti},
  {Glanzman}, {Godfrey}, {Grenier}, {Grondin}, {Grove}, {Guillemot}, {Guiriec},
  {Hadasch}, {Harding}, {Hays}, {Hewitt}, {Hill}, {Horan}, {Iafrate}, {Jogler},
  {J{\'o}hannesson}, {Johnson}, {Johnson}, {Johnson}, {Johnson}, {Kamae},
  {Kataoka}, {Katsuta}, {Kuss}, {La Mura}, {Landriu}, {Larsson}, {Latronico},
  {Lemoine-Goumard}, {Li}, {Li}, {Longo}, {Loparco}, {Lott}, {Lovellette},
  {Lubrano}, {Madejski}, {Massaro}, {Mayer}, {Mazziotta}, {McEnery},
  {Michelson}, {Mirabal}, {Mizuno}, {Moiseev}, {Mongelli}, {Monzani},
  {Morselli}, {Moskalenko}, {Murgia}, {Nuss}, {Ohno}, {Ohsugi}, {Omodei},
  {Orienti}, {Orlando}, {Ormes}, {Paneque}, {Panetta}, {Perkins},
  {Pesce-Rollins}, {Piron}, {Pivato}, {Porter}, {Racusin}, {Rando}, {Razzano},
  {Razzaque}, {Reimer}, {Reimer}, {Reposeur}, {Rochester}, {Romani},
  {Salvetti}, {S{\'a}nchez-Conde}, {Saz Parkinson}, {Schulz}, {Siskind},
  {Smith}, {Spada}, {Spandre}, {Spinelli}, {Stephens}, {Strong}, {Suson},
  {Takahashi}, {Takahashi}, {Tanaka}, {Thayer}, {Thayer}, {Thompson},
  {Tibaldo}, {Tibolla}, {Torres}, {Torresi}, {Tosti}, {Troja}, {Van Klaveren},
  {Vianello}, {Winer}, {Wood}, {Wood}, {Zimmer}, \& {Fermi-LAT
  Collaboration}}]{2015ApJS..218...23A}
{Acero}, F., {Ackermann}, M., {Ajello}, M., {et~al.} 2015, \apjs, 218, 23,
  \dodoi{10.1088/0067-0049/218/2/23}

\bibitem[{{Andersen} \& {Ransom}(2018)}]{Andersen2018}
{Andersen}, B.~C., \& {Ransom}, S.~M. 2018, \apjl, 863, L13,
  \dodoi{10.3847/2041-8213/aad59f}

\bibitem[{{Astropy Collaboration} {et~al.}(2013){Astropy Collaboration},
  {Robitaille}, {Tollerud}, {Greenfield}, {Droettboom}, {Bray}, {Aldcroft},
  {Davis}, {Ginsburg}, {Price-Whelan}, {Kerzendorf}, {Conley}, {Crighton},
  {Barbary}, {Muna}, {Ferguson}, {Grollier}, {Parikh}, {Nair}, {Unther},
  {Deil}, {Woillez}, {Conseil}, {Kramer}, {Turner}, {Singer}, {Fox}, {Weaver},
  {Zabalza}, {Edwards}, {Azalee Bostroem}, {Burke}, {Casey}, {Crawford},
  {Dencheva}, {Ely}, {Jenness}, {Labrie}, {Lim}, {Pierfederici}, {Pontzen},
  {Ptak}, {Refsdal}, {Servillat}, \& {Streicher}}]{2013A&A...558A..33A}
{Astropy Collaboration}, {Robitaille}, T.~P., {Tollerud}, E.~J., {et~al.} 2013,
  \aap, 558, A33, \dodoi{10.1051/0004-6361/201322068}

\bibitem[{{Astropy Collaboration} {et~al.}(2018){Astropy Collaboration},
  {Price-Whelan}, {Sip{\H{o}}cz}, {G{\"u}nther}, {Lim}, {Crawford}, {Conseil},
  {Shupe}, {Craig}, {Dencheva}, {Ginsburg}, {VanderPlas}, {Bradley},
  {P{\'e}rez-Su{\'a}rez}, {de Val-Borro}, {Aldcroft}, {Cruz}, {Robitaille},
  {Tollerud}, {Ardelean}, {Babej}, {Bach}, {Bachetti}, {Bakanov}, {Bamford},
  {Barentsen}, {Barmby}, {Baumbach}, {Berry}, {Biscani}, {Boquien}, {Bostroem},
  {Bouma}, {Brammer}, {Bray}, {Breytenbach}, {Buddelmeijer}, {Burke},
  {Calderone}, {Cano Rodr{\'\i}guez}, {Cara}, {Cardoso}, {Cheedella}, {Copin},
  {Corrales}, {Crichton}, {D'Avella}, {Deil}, {Depagne}, {Dietrich}, {Donath},
  {Droettboom}, {Earl}, {Erben}, {Fabbro}, {Ferreira}, {Finethy}, {Fox},
  {Garrison}, {Gibbons}, {Goldstein}, {Gommers}, {Greco}, {Greenfield},
  {Groener}, {Grollier}, {Hagen}, {Hirst}, {Homeier}, {Horton}, {Hosseinzadeh},
  {Hu}, {Hunkeler}, {Ivezi{\'c}}, {Jain}, {Jenness}, {Kanarek}, {Kendrew},
  {Kern}, {Kerzendorf}, {Khvalko}, {King}, {Kirkby}, {Kulkarni}, {Kumar},
  {Lee}, {Lenz}, {Littlefair}, {Ma}, {Macleod}, {Mastropietro}, {McCully},
  {Montagnac}, {Morris}, {Mueller}, {Mumford}, {Muna}, {Murphy}, {Nelson},
  {Nguyen}, {Ninan}, {N{\"o}the}, {Ogaz}, {Oh}, {Parejko}, {Parley}, {Pascual},
  {Patil}, {Patil}, {Plunkett}, {Prochaska}, {Rastogi}, {Reddy Janga},
  {Sabater}, {Sakurikar}, {Seifert}, {Sherbert}, {Sherwood-Taylor}, {Shih},
  {Sick}, {Silbiger}, {Singanamalla}, {Singer}, {Sladen}, {Sooley},
  {Sornarajah}, {Streicher}, {Teuben}, {Thomas}, {Tremblay}, {Turner},
  {Terr{\'o}n}, {van Kerkwijk}, {de la Vega}, {Watkins}, {Weaver}, {Whitmore},
  {Woillez}, {Zabalza}, \& {Astropy Contributors}}]{2018AJ....156..123A}
{Astropy Collaboration}, {Price-Whelan}, A.~M., {Sip{\H{o}}cz}, B.~M., {et~al.}
  2018, \aj, 156, 123, \dodoi{10.3847/1538-3881/aabc4f}

\bibitem[{{Backer}(1970)}]{1970Natur.228...42B}
{Backer}, D.~C. 1970, \nat, 228, 42, \dodoi{10.1038/228042a0}

\bibitem[{{Backer} {et~al.}(1982){Backer}, {Kulkarni}, {Heiles}, {Davis}, \&
  {Goss}}]{1982Natur.300..615B}
{Backer}, D.~C., {Kulkarni}, S.~R., {Heiles}, C., {Davis}, M.~M., \& {Goss},
  W.~M. 1982, \nat, 300, 615, \dodoi{10.1038/300615a0}

\bibitem[{{Barrett} {et~al.}(2020){Barrett}, {Dieck}, {Beasley}, {Mason}, \&
  {Singh}}]{2020AdSpR..66.1226B}
{Barrett}, P., {Dieck}, C., {Beasley}, A.~J., {Mason}, P.~A., \& {Singh}, K.~P.
  2020, Advances in Space Research, 66, 1226, \dodoi{10.1016/j.asr.2020.04.007}

\bibitem[{{Bates} {et~al.}(2013){Bates}, {Lorimer}, \&
  {Verbiest}}]{2013MNRAS.431.1352B}
{Bates}, S.~D., {Lorimer}, D.~R., \& {Verbiest}, J.~P.~W. 2013, \mnras, 431,
  1352, \dodoi{10.1093/mnras/stt257}

\bibitem[{{Bell} {et~al.}(2016){Bell}, {Murphy}, {Johnston}, {Kaplan}, {Croft},
  {Hancock}, {Callingham}, {Zic}, {Dobie}, {Swiggum}, {Rowlinson},
  {Hurley-Walker}, {Offringa}, {Bernardi}, {Bowman}, {Briggs}, {Cappallo},
  {Deshpande}, {Gaensler}, {Greenhill}, {Hazelton}, {Johnston-Hollitt},
  {Lonsdale}, {McWhirter}, {Mitchell}, {Morales}, {Morgan}, {Oberoi}, {Ord},
  {Prabu}, {Shankar}, {Srivani}, {Subrahmanyan}, {Tingay}, {Wayth}, {Webster},
  {Williams}, \& {Williams}}]{2016MNRAS.461..908B}
{Bell}, M.~E., {Murphy}, T., {Johnston}, S., {et~al.} 2016, \mnras, 461, 908,
  \dodoi{10.1093/mnras/stw1293}

\bibitem[{{Bhakta} {et~al.}(2017){Bhakta}, {Deneva}, {Frail}, {de Gasperin},
  {Intema}, {Jagannathan}, \& {Mooley}}]{Bhakta2017}
{Bhakta}, D., {Deneva}, J.~S., {Frail}, D.~A., {et~al.} 2017, \mnras, 468,
  2526, \dodoi{10.1093/mnras/stx656}

\bibitem[{{Bhat} {et~al.}(2004){Bhat}, {Cordes}, {Camilo}, {Nice}, \&
  {Lorimer}}]{2004ApJ...605..759B}
{Bhat}, N.~D.~R., {Cordes}, J.~M., {Camilo}, F., {Nice}, D.~J., \& {Lorimer},
  D.~R. 2004, \apj, 605, 759, \dodoi{10.1086/382680}

\bibitem[{{Bhat} {et~al.}(2011){Bhat}, {Cordes}, {Cox}, {Deneva}, {Hankins},
  {Lazio}, \& {McLaughlin}}]{Bhat2011}
{Bhat}, N.~D.~R., {Cordes}, J.~M., {Cox}, P.~J., {et~al.} 2011, \apj, 732, 14,
  \dodoi{10.1088/0004-637X/732/1/14}

\bibitem[{{Boller} {et~al.}(2016){Boller}, {Freyberg}, {Tr{\"u}mper}, {Haberl},
  {Voges}, \& {Nandra}}]{2016A&A...588A.103B}
{Boller}, T., {Freyberg}, M.~J., {Tr{\"u}mper}, J., {et~al.} 2016, \aap, 588,
  A103, \dodoi{10.1051/0004-6361/201525648}

\bibitem[{{Camilo} {et~al.}(2000){Camilo}, {Lorimer}, {Freire}, {Lyne}, \&
  {Manchester}}]{2000ApJ...535..975C}
{Camilo}, F., {Lorimer}, D.~R., {Freire}, P., {Lyne}, A.~G., \& {Manchester},
  R.~N. 2000, \apj, 535, 975, \dodoi{10.1086/308859}

\bibitem[{{Cioni} {et~al.}(2011){Cioni}, {Clementini}, {Girardi}, {Guandalini},
  {Gullieuszik}, {Miszalski}, {Moretti}, {Ripepi}, {Rubele}, {Bagheri},
  {Bekki}, {Cross}, {de Blok}, {de Grijs}, {Emerson}, {Evans}, {Gibson},
  {Gonzales-Solares}, {Groenewegen}, {Irwin}, {Ivanov}, {Lewis}, {Marconi},
  {Marquette}, {Mastropietro}, {Moore}, {Napiwotzki}, {Naylor}, {Oliveira},
  {Read}, {Sutorius}, {van Loon}, {Wilkinson}, \& {Wood}}]{2011AA...527A.116C}
{Cioni}, M. R.~L., {Clementini}, G., {Girardi}, L., {et~al.} 2011, \aap, 527,
  A116, \dodoi{10.1051/0004-6361/201016137}

\bibitem[{{Cordes} \& {Lazio}(2002)}]{2002astro.ph..7156C}
{Cordes}, J.~M., \& {Lazio}, T.~J.~W. 2002, arXiv e-prints, astro.
\newblock \doarXiv{astro-ph/0207156}

\bibitem[{{Cordes} \& {Shannon}(2008)}]{Cordes2008}
{Cordes}, J.~M., \& {Shannon}, R.~M. 2008, \apj, 682, 1152,
  \dodoi{10.1086/589425}

\bibitem[{{Crawford} {et~al.}(2001){Crawford}, {Kaspi}, {Manchester}, {Lyne},
  {Camilo}, \& {D'Amico}}]{Crawford2001}
{Crawford}, F., {Kaspi}, V.~M., {Manchester}, R.~N., {et~al.} 2001, \apj, 553,
  367, \dodoi{10.1086/320635}

\bibitem[{{Dai} {et~al.}(2016){Dai}, {Johnston}, {Bell}, {Coles}, {Hobbs},
  {Ekers}, \& {Lenc}}]{2016MNRAS.462.3115D}
{Dai}, S., {Johnston}, S., {Bell}, M.~E., {et~al.} 2016, \mnras, 462, 3115,
  \dodoi{10.1093/mnras/stw1871}

\bibitem[{{Dai} {et~al.}(2015){Dai}, {Hobbs}, {Manchester}, {Kerr}, {Shannon},
  {van Straten}, {Mata}, {Bailes}, {Bhat}, {Burke-Spolaor}, {Coles},
  {Johnston}, {Keith}, {Levin}, {Os{\l}owski}, {Reardon}, {Ravi}, {Sarkissian},
  {Tiburzi}, {Toomey}, {Wang}, {Wang}, {Wen}, {Xu}, {Yan}, \&
  {Zhu}}]{2015MNRAS.449.3223D}
{Dai}, S., {Hobbs}, G., {Manchester}, R.~N., {et~al.} 2015, \mnras, 449, 3223,
  \dodoi{10.1093/mnras/stv508}

\bibitem[{{Dalton} {et~al.}(2006){Dalton}, {Caldwell}, {Ward}, {Whalley},
  {Woodhouse}, {Edeson}, {Clark}, {Beard}, {Gallie}, {Todd}, {Strachan},
  {Bezawada}, {Sutherland}, \& {Emerson}}]{2006SPIE.6269E..0XD}
{Dalton}, G.~B., {Caldwell}, M., {Ward}, A.~K., {et~al.} 2006, in Society of
  Photo-Optical Instrumentation Engineers (SPIE) Conference Series, Vol. 6269,
  Society of Photo-Optical Instrumentation Engineers (SPIE) Conference Series,
  ed. I.~S. {McLean} \& M.~{Iye}, 62690X, \dodoi{10.1117/12.670018}

\bibitem[{{Damico} {et~al.}(1985){Damico}, {Manchester}, {Durdin}, \&
  {Erickson}}]{1985PASA....6..174D}
{Damico}, N., {Manchester}, R.~N., {Durdin}, J.~M., \& {Erickson}, W.~C. 1985,
  \pasa, 6, 174, \dodoi{10.1017/S1323358000018026}

\bibitem[{{Emerson} {et~al.}(2006){Emerson}, {McPherson}, \&
  {Sutherland}}]{2006Msngr.126...41E}
{Emerson}, J., {McPherson}, A., \& {Sutherland}, W. 2006, The Messenger, 126,
  41

\bibitem[{{Frail} {et~al.}(1991){Frail}, {Cordes}, {Hankins}, \&
  {Weisberg}}]{Frail1991}
{Frail}, D.~A., {Cordes}, J.~M., {Hankins}, T.~H., \& {Weisberg}, J.~M. 1991,
  \apj, 382, 168, \dodoi{10.1086/170705}

\bibitem[{{Frail} {et~al.}(2016{\natexlab{a}}){Frail}, {Jagannathan}, {Mooley},
  \& {Intema}}]{2016ApJ...829..119F}
{Frail}, D.~A., {Jagannathan}, P., {Mooley}, K.~P., \& {Intema}, H.~T.
  2016{\natexlab{a}}, \apj, 829, 119, \dodoi{10.3847/0004-637X/829/2/119}

\bibitem[{{Frail} {et~al.}(2016{\natexlab{b}}){Frail}, {Mooley}, {Jagannathan},
  \& {Intema}}]{2016MNRAS.461.1062F}
{Frail}, D.~A., {Mooley}, K.~P., {Jagannathan}, P., \& {Intema}, H.~T.
  2016{\natexlab{b}}, \mnras, 461, 1062, \dodoi{10.1093/mnras/stw1390}

\bibitem[{{Frail} {et~al.}(2018){Frail}, {Ray}, {Mooley}, {Hancock}, {Burnett},
  {Jagannathan}, {Ferrara}, {Intema}, {de Gasperin}, {Demorest}, {Stovall}, \&
  {McKinnon}}]{Frail2018}
{Frail}, D.~A., {Ray}, P.~S., {Mooley}, K.~P., {et~al.} 2018, \mnras, 475, 942,
  \dodoi{10.1093/mnras/stx3281}

\bibitem[{{Gaensler} {et~al.}(2005){Gaensler}, {Haverkorn}, {Staveley-Smith},
  {Dickey}, {McClure-Griffiths}, {Dickel}, \& {Wolleben}}]{Gaensler2005}
{Gaensler}, B.~M., {Haverkorn}, M., {Staveley-Smith}, L., {et~al.} 2005,
  Science, 307, 1610, \dodoi{10.1126/science.1108832}

\bibitem[{{Gaensler} {et~al.}(2010){Gaensler}, {Landecker}, {Taylor}, \&
  {POSSUM Collaboration}}]{Gaensler2010}
{Gaensler}, B.~M., {Landecker}, T.~L., {Taylor}, A.~R., \& {POSSUM
  Collaboration}. 2010, in American Astronomical Society Meeting Abstracts,
  Vol. 215, American Astronomical Society Meeting Abstracts \#215, 470.13

\bibitem[{{Gaensler} {et~al.}(1998){Gaensler}, {Manchester}, \&
  {Green}}]{Gaensler1998}
{Gaensler}, B.~M., {Manchester}, R.~N., \& {Green}, A.~J. 1998, \mnras, 296,
  813, \dodoi{10.1046/j.1365-8711.1998.01387.x}

\bibitem[{{Gaia Collaboration} {et~al.}(2016){Gaia Collaboration}, {Prusti},
  {de Bruijne}, {Brown}, {Vallenari}, {Babusiaux}, {Bailer-Jones}, {Bastian},
  {Biermann}, {Evans}, \& et~al.}]{GaiaCollaboration2016}
{Gaia Collaboration}, {Prusti}, T., {de Bruijne}, J.~H.~J., {et~al.} 2016,
  \aap, 595, A1, \dodoi{10.1051/0004-6361/201629272}

\bibitem[{{Gaia Collaboration} {et~al.}(2018){Gaia Collaboration}, {Brown},
  {Vallenari}, {Prusti}, {de Bruijne}, {Babusiaux}, {Bailer-Jones}, {Biermann},
  {Evans}, {Eyer}, {Jansen}, {Jordi}, {Klioner}, {Lammers}, {Lindegren},
  {Luri}, {Mignard}, {Panem}, {Pourbaix}, {Randich}, {Sartoretti}, {Siddiqui},
  {Soubiran}, {van Leeuwen}, {Walton}, {Arenou}, {Bastian}, {Cropper},
  {Drimmel}, {Katz}, {Lattanzi}, {Bakker}, {Cacciari}, {Casta{\~n}eda},
  {Chaoul}, {Cheek}, {De Angeli}, {Fabricius}, {Guerra}, {Holl}, {Masana},
  {Messineo}, {Mowlavi}, {Nienartowicz}, {Panuzzo}, {Portell}, {Riello},
  {Seabroke}, {Tanga}, {Th{\'e}venin}, {Gracia-Abril}, {Comoretto},
  {Garcia-Reinaldos}, {Teyssier}, {Altmann}, {Andrae}, {Audard},
  {Bellas-Velidis}, {Benson}, {Berthier}, {Blomme}, {Burgess}, {Busso},
  {Carry}, {Cellino}, {Clementini}, {Clotet}, {Creevey}, {Davidson}, {De
  Ridder}, {Delchambre}, {Dell'Oro}, {Ducourant},
  {Fern{\'a}ndez-Hern{\'a}ndez}, {Fouesneau}, {Fr{\'e}mat}, {Galluccio},
  {Garc{\'\i}a-Torres}, {Gonz{\'a}lez-N{\'u}{\~n}ez}, {Gonz{\'a}lez-Vidal},
  {Gosset}, {Guy}, {Halbwachs}, {Hambly}, {Harrison}, {Hern{\'a}ndez},
  {Hestroffer}, {Hodgkin}, {Hutton}, {Jasniewicz}, {Jean-Antoine-Piccolo},
  {Jordan}, {Korn}, {Krone-Martins}, {Lanzafame}, {Lebzelter}, {L{\"o}ffler},
  {Manteiga}, {Marrese}, {Mart{\'\i}n-Fleitas}, {Moitinho}, {Mora}, {Muinonen},
  {Osinde}, {Pancino}, {Pauwels}, {Petit}, {Recio-Blanco}, {Richards},
  {Rimoldini}, {Robin}, {Sarro}, {Siopis}, {Smith}, {Sozzetti}, {S{\"u}veges},
  {Torra}, {van Reeven}, {Abbas}, {Abreu Aramburu}, {Accart}, {Aerts},
  {Altavilla}, {{\'A}lvarez}, {Alvarez}, {Alves}, {Anderson}, {Andrei},
  {Anglada Varela}, {Antiche}, {Antoja}, {Arcay}, {Astraatmadja}, {Bach},
  {Baker}, {Balaguer-N{\'u}{\~n}ez}, {Balm}, {Barache}, {Barata}, {Barbato},
  {Barblan}, {Barklem}, {Barrado}, {Barros}, {Barstow}, {Bartholom{\'e}
  Mu{\~n}oz}, {Bassilana}, {Becciani}, {Bellazzini}, {Berihuete}, {Bertone},
  {Bianchi}, {Bienaym{\'e}}, {Blanco-Cuaresma}, {Boch}, {Boeche}, {Bombrun},
  {Borrachero}, {Bossini}, {Bouquillon}, {Bourda}, {Bragaglia}, {Bramante},
  {Breddels}, {Bressan}, {Brouillet}, {Br{\"u}semeister}, {Brugaletta},
  {Bucciarelli}, {Burlacu}, {Busonero}, {Butkevich}, {Buzzi}, {Caffau},
  {Cancelliere}, {Cannizzaro}, {Cantat-Gaudin}, {Carballo}, {Carlucci},
  {Carrasco}, {Casamiquela}, {Castellani}, {Castro-Ginard}, {Charlot},
  {Chemin}, {Chiavassa}, {Cocozza}, {Costigan}, {Cowell}, {Crifo}, {Crosta},
  {Crowley}, {Cuypers}, {Dafonte}, {Damerdji}, {Dapergolas}, {David}, {David},
  {de Laverny}, {De Luise}, {De March}, {de Martino}, {de Souza}, {de Torres},
  {Debosscher}, {del Pozo}, {Delbo}, {Delgado}, {Delgado}, {Di Matteo},
  {Diakite}, {Diener}, {Distefano}, {Dolding}, {Drazinos}, {Dur{\'a}n},
  {Edvardsson}, {Enke}, {Eriksson}, {Esquej}, {Eynard Bontemps}, {Fabre},
  {Fabrizio}, {Faigler}, {Falc{\~a}o}, {Farr{\`a}s Casas}, {Federici},
  {Fedorets}, {Fernique}, {Figueras}, {Filippi}, {Findeisen}, {Fonti},
  {Fraile}, {Fraser}, {Fr{\'e}zouls}, {Gai}, {Galleti}, {Garabato},
  {Garc{\'\i}a-Sedano}, {Garofalo}, {Garralda}, {Gavel}, {Gavras}, {Gerssen},
  {Geyer}, {Giacobbe}, {Gilmore}, {Girona}, {Giuffrida}, {Glass}, {Gomes},
  {Granvik}, {Gueguen}, {Guerrier}, {Guiraud}, {Guti{\'e}rrez-S{\'a}nchez},
  {Haigron}, {Hatzidimitriou}, {Hauser}, {Haywood}, {Heiter}, {Helmi}, {Heu},
  {Hilger}, {Hobbs}, {Hofmann}, {Holland}, {Huckle}, {Hypki}, {Icardi},
  {Jan{\ss}en}, {Jevardat de Fombelle}, {Jonker}, {Juh{\'a}sz}, {Julbe},
  {Karampelas}, {Kewley}, {Klar}, {Kochoska}, {Kohley}, {Kolenberg},
  {Kontizas}, {Kontizas}, {Koposov}, {Kordopatis}, {Kostrzewa-Rutkowska},
  {Koubsky}, {Lambert}, {Lanza}, {Lasne}, {Lavigne}, {Le Fustec}, {Le
  Poncin-Lafitte}, {Lebreton}, {Leccia}, {Leclerc}, {Lecoeur-Taibi},
  {Lenhardt}, {Leroux}, {Liao}, {Licata}, {Lindstr{\o}m}, {Lister}, {Livanou},
  {Lobel}, {L{\'o}pez}, {Managau}, {Mann}, {Mantelet}, {Marchal}, {Marchant},
  {Marconi}, {Marinoni}, {Marschalk{\'o}}, {Marshall}, {Martino}, {Marton},
  {Mary}, {Massari}, {Matijevi{\v{c}}}, {Mazeh}, {McMillan}, {Messina},
  {Michalik}, {Millar}, {Molina}, {Molinaro}, {Moln{\'a}r}, {Montegriffo},
  {Mor}, {Morbidelli}, {Morel}, {Morris}, {Mulone}, {Muraveva}, {Musella},
  {Nelemans}, {Nicastro}, {Noval}, {O'Mullane}, {Ord{\'e}novic},
  {Ord{\'o}{\~n}ez-Blanco}, {Osborne}, {Pagani}, {Pagano}, {Pailler},
  {Palacin}, {Palaversa}, {Panahi}, {Pawlak}, {Piersimoni}, {Pineau}, {Plachy},
  {Plum}, {Poggio}, {Poujoulet}, {Pr{\v{s}}a}, {Pulone}, {Racero}, {Ragaini},
  {Rambaux}, {Ramos-Lerate}, {Regibo}, {Reyl{\'e}}, {Riclet}, {Ripepi}, {Riva},
  {Rivard}, {Rixon}, {Roegiers}, {Roelens}, {Romero-G{\'o}mez}, {Rowell},
  {Royer}, {Ruiz-Dern}, {Sadowski}, {Sagrist{\`a} Sell{\'e}s}, {Sahlmann},
  {Salgado}, {Salguero}, {Sanna}, {Santana-Ros}, {Sarasso}, {Savietto},
  {Schultheis}, {Sciacca}, {Segol}, {Segovia}, {S{\'e}gransan}, {Shih},
  {Siltala}, {Silva}, {Smart}, {Smith}, {Solano}, {Solitro}, {Sordo}, {Soria
  Nieto}, {Souchay}, {Spagna}, {Spoto}, {Stampa}, {Steele},
  {Steidelm{\"u}ller}, {Stephenson}, {Stoev}, {Suess}, {Surdej}, {Szabados},
  {Szegedi-Elek}, {Tapiador}, {Taris}, {Tauran}, {Taylor}, {Teixeira},
  {Terrett}, {Teyssand ier}, {Thuillot}, {Titarenko}, {Torra Clotet}, {Turon},
  {Ulla}, {Utrilla}, {Uzzi}, {Vaillant}, {Valentini}, {Valette}, {van Elteren},
  {Van Hemelryck}, {van Leeuwen}, {Vaschetto}, {Vecchiato}, {Veljanoski},
  {Viala}, {Vicente}, {Vogt}, {von Essen}, {Voss}, {Votruba}, {Voutsinas},
  {Walmsley}, {Weiler}, {Wertz}, {Wevers}, {Wyrzykowski}, {Yoldas},
  {{\v{Z}}erjal}, {Ziaeepour}, {Zorec}, {Zschocke}, {Zucker}, {Zurbach}, \&
  {Zwitter}}]{2018A&A...616A...1G}
{Gaia Collaboration}, {Brown}, A.~G.~A., {Vallenari}, A., {et~al.} 2018, \aap,
  616, A1, \dodoi{10.1051/0004-6361/201833051}

\bibitem[{{Han} {et~al.}(1998){Han}, {Manchester}, {Xu}, \& {Qiao}}]{Han1998}
{Han}, J.~L., {Manchester}, R.~N., {Xu}, R.~X., \& {Qiao}, G.~J. 1998, \mnras,
  300, 373, \dodoi{10.1046/j.1365-8711.1998.01869.x}

\bibitem[{Harris {et~al.}(2020)Harris, Millman, van~der Walt, Gommers,
  Virtanen, Cournapeau, Wieser, Taylor, Berg, Smith, Kern, Picus, Hoyer, van
  Kerkwijk, Brett, Haldane, del R{\'{i}}o, Wiebe, Peterson,
  G{\'{e}}rard-Marchant, Sheppard, Reddy, Weckesser, Abbasi, Gohlke, \&
  Oliphant}]{harris2020array}
Harris, C.~R., Millman, K.~J., van~der Walt, S.~J., {et~al.} 2020, \nat, 585,
  357, \dodoi{10.1038/s41586-020-2649-2}

\bibitem[{{Hewish} {et~al.}(1968){Hewish}, {Bell}, {Pilkington}, {Scott}, \&
  {Collins}}]{1968Natur.217..709H}
{Hewish}, A., {Bell}, S.~J., {Pilkington}, J.~D.~H., {Scott}, P.~F., \&
  {Collins}, R.~A. 1968, \nat, 217, 709, \dodoi{10.1038/217709a0}

\bibitem[{{Heywood}(2020)}]{2020ascl.soft09003H}
{Heywood}, I. 2020, {oxkat: Semi-automated imaging of MeerKAT observations}.
\newblock \doeprint{2009.003}

\bibitem[{{Hobbs} {et~al.}(2016){Hobbs}, {Heywood}, {Bell}, {Kerr},
  {Rowlinson}, {Johnston}, {Shannon}, {Voronkov}, {Ward}, {Banyer}, {Hancock},
  {Murphy}, {Allison}, {Amy}, {Ball}, {Bannister}, {Bock}, {Brodrick},
  {Brothers}, {Brown}, {Bunton}, {Chapman}, {Chippendale}, {Chung}, {DeBoer},
  {Diamond}, {Edwards}, {Ekers}, {Ferris}, {Forsyth}, {Gough}, {Grancea},
  {Gupta}, {Harvey-Smith}, {Hay}, {Hayman}, {Hotan}, {Hoyle}, {Humphreys},
  {Indermuehle}, {Jacka}, {Jackson}, {Jackson}, {Jeganathan}, {Joseph},
  {Kendall}, {Kiraly}, {Koribalski}, {Leach}, {Lenc}, {MacLeod}, {Mader},
  {Marquarding}, {Marvil}, {McClure-Griffiths}, {McConnell}, {Mirtschin},
  {Neuhold}, {Ng}, {Norris}, {O'Sullivan}, {Pearce}, {Phillips}, {Popping},
  {Qiao}, {Reynolds}, {Roberts}, {Sault}, {Schinckel}, {Serra}, {Shaw},
  {Shimwell}, {Storey}, {Sweetnam}, {Tzioumis}, {Westmeier}, {Whiting}, \&
  {Wilson}}]{Hobbs2016}
{Hobbs}, G., {Heywood}, I., {Bell}, M.~E., {et~al.} 2016, \mnras, 456, 3948,
  \dodoi{10.1093/mnras/stv2893}

\bibitem[{{Hobbs} {et~al.}(2020){Hobbs}, {Manchester}, {Dunning}, {Jameson},
  {Roberts}, {George}, {Green}, {Tuthill}, {Toomey}, {Kaczmarek}, {Mader},
  {Marquarding}, {Ahmed}, {Amy}, {Bailes}, {Beresford}, {Bhat}, {Bock},
  {Bourne}, {Bowen}, {Brothers}, {Cameron}, {Carretti}, {Carter}, {Castillo},
  {Chekkala}, {Cheng}, {Chung}, {Craig}, {Dai}, {Dawson}, {Dempsey}, {Doherty},
  {Dong}, {Edwards}, {Ergesh}, {Gao}, {Han}, {Hayman}, {Indermuehle},
  {Jeganathan}, {Johnston}, {Kanoniuk}, {Kesteven}, {Kramer}, {Leach},
  {Mcintyre}, {Moss}, {Os{\l}owski}, {Phillips}, {Pope}, {Preisig}, {Price},
  {Reeves}, {Reilly}, {Reynolds}, {Robishaw}, {Roush}, {Ruckley}, {Sadler},
  {Sarkissian}, {Severs}, {Shannon}, {Smart}, {Smith}, {Smith}, {Sobey},
  {Staveley-Smith}, {Tzioumis}, {van Straten}, {Wang}, {Wen}, \&
  {Whiting}}]{2020PASA...37...12H}
{Hobbs}, G., {Manchester}, R.~N., {Dunning}, A., {et~al.} 2020, \pasa, 37,
  e012, \dodoi{10.1017/pasa.2020.2}

\bibitem[{{Hotan} {et~al.}(2004){Hotan}, {van Straten}, \&
  {Manchester}}]{2004PASA...21..302H}
{Hotan}, A.~W., {van Straten}, W., \& {Manchester}, R.~N. 2004, \pasa, 21, 302,
  \dodoi{10.1071/AS04022}

\bibitem[{{Hotan} {et~al.}(2021){Hotan}, {Bunton}, {Chippendale}, {Whiting},
  {Tuthill}, {Moss}, {McConnell}, {Amy}, {Huynh}, {Allison}, {Anderson},
  {Bannister}, {Bastholm}, {Beresford}, {Bock}, {Bolton}, {Chapman}, {Chow},
  {Collier}, {Cooray}, {Cornwell}, {Diamond}, {Edwards}, {Feain}, {Franzen},
  {George}, {Gupta}, {Hampson}, {Harvey-Smith}, {Hayman}, {Heywood}, {Jacka},
  {Jackson}, {Jackson}, {Jeganathan}, {Johnston}, {Kesteven}, {Kleiner},
  {Koribalski}, {Lee-Waddell}, {Lenc}, {Lensson}, {Mackay}, {Mahony},
  {McClure-Griffiths}, {McConigley}, {Mirtschin}, {Ng}, {Norris}, {Pearce},
  {Phillips}, {Pilawa}, {Raja}, {Reynolds}, {Roberts}, {Roxby}, {Sadler},
  {Shields}, {Schinckel}, {Serra}, {Shaw}, {Sweetnam}, {Troup}, {Tzioumis},
  {Voronkov}, \& {Westmeier}}]{Hotan2021}
{Hotan}, A.~W., {Bunton}, J.~D., {Chippendale}, A.~P., {et~al.} 2021, \pasa,
  38, e009, \dodoi{10.1017/pasa.2021.1}

\bibitem[{{Hunter}(2007)}]{Hunter2007}
{Hunter}, J.~D. 2007, Computing in Science \& Engineering, 9, 90,
  \dodoi{10.1109/MCSE.2007.55}

\bibitem[{{Hurley-Walker} {et~al.}(2017){Hurley-Walker}, {Callingham},
  {Hancock}, {Franzen}, {Hindson}, {Kapi{\'n}ska}, {Morgan}, {Offringa},
  {Wayth}, {Wu}, {Zheng}, {Murphy}, {Bell}, {Dwarakanath}, {For}, {Gaensler},
  {Johnston-Hollitt}, {Lenc}, {Procopio}, {Staveley-Smith}, {Ekers}, {Bowman},
  {Briggs}, {Cappallo}, {Deshpande}, {Greenhill}, {Hazelton}, {Kaplan},
  {Lonsdale}, {McWhirter}, {Mitchell}, {Morales}, {Morgan}, {Oberoi}, {Ord},
  {Prabu}, {Shankar}, {Srivani}, {Subrahmanyan}, {Tingay}, {Webster},
  {Williams}, \& {Williams}}]{Hurley-Walker2017}
{Hurley-Walker}, N., {Callingham}, J.~R., {Hancock}, P.~J., {et~al.} 2017,
  \mnras, 464, 1146, \dodoi{10.1093/mnras/stw2337}

\bibitem[{{Hyman} {et~al.}(2019){Hyman}, {Frail}, {Deneva}, {Kassim},
  {McLaughlin}, {Kooi}, {Ray}, \& {Polisensky}}]{2019ApJ...876...20H}
{Hyman}, S.~D., {Frail}, D.~A., {Deneva}, J.~S., {et~al.} 2019, \apj, 876, 20,
  \dodoi{10.3847/1538-4357/ab11c8}

\bibitem[{{Hyman} {et~al.}(2002){Hyman}, {Lazio}, {Kassim}, \&
  {Bartleson}}]{2002AJ....123.1497H}
{Hyman}, S.~D., {Lazio}, T. J.~W., {Kassim}, N.~E., \& {Bartleson}, A.~L. 2002,
  \aj, 123, 1497, \dodoi{10.1086/338905}

\bibitem[{{Hyman} {et~al.}(2005){Hyman}, {Lazio}, {Kassim}, {Ray}, {Markwardt},
  \& {Yusef-Zadeh}}]{2005Natur.434...50H}
{Hyman}, S.~D., {Lazio}, T. J.~W., {Kassim}, N.~E., {et~al.} 2005, \nat, 434,
  50, \dodoi{10.1038/nature03400}

\bibitem[{{Hyman} {et~al.}(2021){Hyman}, {Frail}, {Deneva}, {Kassim},
  {Giacintucci}, {Kooi}, {Lazio}, {Joyner}, {Peters}, {Gajjar}, \&
  {Siemion}}]{2021MNRAS.507.3888H}
{Hyman}, S.~D., {Frail}, D.~A., {Deneva}, J.~S., {et~al.} 2021, \mnras, 507,
  3888, \dodoi{10.1093/mnras/stab1979}

\bibitem[{{Intema} {et~al.}(2017){Intema}, {Jagannathan}, {Mooley}, \&
  {Frail}}]{Intema2017}
{Intema}, H.~T., {Jagannathan}, P., {Mooley}, K.~P., \& {Frail}, D.~A. 2017,
  \aap, 598, A78, \dodoi{10.1051/0004-6361/201628536}

\bibitem[{{Jankowski} {et~al.}(2018){Jankowski}, {van Straten}, {Keane},
  {Bailes}, {Barr}, {Johnston}, \& {Kerr}}]{2018MNRAS.473.4436J}
{Jankowski}, F., {van Straten}, W., {Keane}, E.~F., {et~al.} 2018, \mnras, 473,
  4436, \dodoi{10.1093/mnras/stx2476}

\bibitem[{{Johnston} \& {Kerr}(2018)}]{Johnston2018}
{Johnston}, S., \& {Kerr}, M. 2018, \mnras, 474, 4629,
  \dodoi{10.1093/mnras/stx3095}

\bibitem[{{Johnston} {et~al.}(2022){Johnston}, {Parthasarathy}, {Main},
  {Ridley}, {Koribalski}, {Bailes}, {Buchner}, {Geyer}, {Karastergiou},
  {Keith}, {Kramer}, {Serylak}, {Shannon}, {Spiewak}, \&
  {Krishnan}}]{Johnston2022}
{Johnston}, S., {Parthasarathy}, A., {Main}, R.~A., {et~al.} 2022, \mnras, 509,
  5209, \dodoi{10.1093/mnras/stab3360}

\bibitem[{{Kao} {et~al.}(2016){Kao}, {Hallinan}, {Pineda}, {Escala},
  {Burgasser}, {Bourke}, \& {Stevenson}}]{2016ApJ...818...24K}
{Kao}, M.~M., {Hallinan}, G., {Pineda}, J.~S., {et~al.} 2016, \apj, 818, 24,
  \dodoi{10.3847/0004-637X/818/1/24}

\bibitem[{{Kaplan} {et~al.}(2019){Kaplan}, {Dai}, {Lenc}, {Zic}, {Swiggum},
  {Murphy}, {Anderson}, {Cameron}, {Dobie}, {Hobbs}, {Kaczmarek}, {Lynch}, \&
  {Toomey}}]{2019ApJ...884...96K}
{Kaplan}, D.~L., {Dai}, S., {Lenc}, E., {et~al.} 2019, \apj, 884, 96,
  \dodoi{10.3847/1538-4357/ab397f}

\bibitem[{{Kenyon} {et~al.}(2018){Kenyon}, {Smirnov}, {Grobler}, \&
  {Perkins}}]{2018MNRAS.478.2399K}
{Kenyon}, J.~S., {Smirnov}, O.~M., {Grobler}, T.~L., \& {Perkins}, S.~J. 2018,
  \mnras, 478, 2399, \dodoi{10.1093/mnras/sty1221}

\bibitem[{{Komesaroff} {et~al.}(1973){Komesaroff}, {Ables}, {Cooke},
  {Hamilton}, \& {McCulloch}}]{Komesaroff1973}
{Komesaroff}, M.~M., {Ables}, J.~G., {Cooke}, D.~J., {Hamilton}, P.~A., \&
  {McCulloch}, P.~M. 1973, \aplett, 15, 169

\bibitem[{{Kramer} {et~al.}(1996){Kramer}, {Xilouris}, {Jessner},
  {Wielebinski}, \& {Timofeev}}]{Kramer1996}
{Kramer}, M., {Xilouris}, K.~M., {Jessner}, A., {Wielebinski}, R., \&
  {Timofeev}, M. 1996, \aap, 306, 867

\bibitem[{{Lacy} {et~al.}(2020){Lacy}, {Baum}, {Chandler}, {Chatterjee},
  {Clarke}, {Deustua}, {English}, {Farnes}, {Gaensler}, {Gugliucci},
  {Hallinan}, {Kent}, {Kimball}, {Law}, {Lazio}, {Marvil}, {Mao}, {Medlin},
  {Mooley}, {Murphy}, {Myers}, {Osten}, {Richards}, {Rosolowsky}, {Rudnick},
  {Schinzel}, {Sivakoff}, {Sjouwerman}, {Taylor}, {White}, {Wrobel},
  {Andernach}, {Beasley}, {Berger}, {Bhatnager}, {Birkinshaw}, {Bower},
  {Brandt}, {Brown}, {Burke-Spolaor}, {Butler}, {Comerford}, {Demorest}, {Fu},
  {Giacintucci}, {Golap}, {G{\"u}th}, {Hales}, {Hiriart}, {Hodge}, {Horesh},
  {Ivezi{\'c}}, {Jarvis}, {Kamble}, {Kassim}, {Liu}, {Loinard}, {Lyons},
  {Masters}, {Mezcua}, {Moellenbrock}, {Mroczkowski}, {Nyland}, {O'Dea},
  {O'Sullivan}, {Peters}, {Radford}, {Rao}, {Robnett}, {Salcido}, {Shen},
  {Sobotka}, {Witz}, {Vaccari}, {van Weeren}, {Vargas}, {Williams}, \&
  {Yoon}}]{Lacy2020}
{Lacy}, M., {Baum}, S.~A., {Chandler}, C.~J., {et~al.} 2020, \pasp, 132,
  035001, \dodoi{10.1088/1538-3873/ab63eb}

\bibitem[{{Lam}(2017)}]{2017ascl.soft06011L}
{Lam}, M.~T. 2017, {PyPulse: PSRFITS handler}.
\newblock \doeprint{1706.011}

\bibitem[{{Lenc} {et~al.}(2018){Lenc}, {Murphy}, {Lynch}, {Kaplan}, \&
  {Zhang}}]{2018MNRAS.478.2835L}
{Lenc}, E., {Murphy}, T., {Lynch}, C.~R., {Kaplan}, D.~L., \& {Zhang}, S.~N.
  2018, \mnras, 478, 2835, \dodoi{10.1093/mnras/sty1304}

\bibitem[{{Lorimer}(2004)}]{Lorimer2004}
{Lorimer}, D.~R. 2004, in Young Neutron Stars and Their Environments, ed.
  F.~{Camilo} \& B.~M. {Gaensler}, Vol. 218, 105.
\newblock \doarXiv{astro-ph/0308501}

\bibitem[{{Lorimer} \& {Kramer}(2012)}]{2012hpa..book.....L}
{Lorimer}, D.~R., \& {Kramer}, M. 2012, {Handbook of Pulsar Astronomy}

\bibitem[{{Lorimer} {et~al.}(2006){Lorimer}, {Faulkner}, {Lyne}, {Manchester},
  {Kramer}, {McLaughlin}, {Hobbs}, {Possenti}, {Stairs}, {Camilo}, {Burgay},
  {D'Amico}, {Corongiu}, \& {Crawford}}]{Lorimer2006}
{Lorimer}, D.~R., {Faulkner}, A.~J., {Lyne}, A.~G., {et~al.} 2006, \mnras, 372,
  777, \dodoi{10.1111/j.1365-2966.2006.10887.x}

\bibitem[{{Lyne}(2009)}]{2009ASSL..357...67L}
{Lyne}, A.~G. 2009, {Intermittent Pulsars}, ed. W.~{Becker}, Vol. 357, 67,
  \dodoi{10.1007/978-3-540-76965-1\_4}

\bibitem[{{Maan} {et~al.}(2018){Maan}, {Bassa}, {van Leeuwen}, {Krishnakumar},
  \& {Joshi}}]{2018ApJ...864...16M}
{Maan}, Y., {Bassa}, C., {van Leeuwen}, J., {Krishnakumar}, M.~A., \& {Joshi},
  B.~C. 2018, \apj, 864, 16, \dodoi{10.3847/1538-4357/aad4ad}

\bibitem[{{Manchester} {et~al.}(2006){Manchester}, {Fan}, {Lyne}, {Kaspi}, \&
  {Crawford}}]{2006ApJ...649..235M}
{Manchester}, R.~N., {Fan}, G., {Lyne}, A.~G., {Kaspi}, V.~M., \& {Crawford},
  F. 2006, \apj, 649, 235, \dodoi{10.1086/505461}

\bibitem[{{Manchester} {et~al.}(2005){Manchester}, {Hobbs}, {Teoh}, \&
  {Hobbs}}]{2005AJ....129.1993M}
{Manchester}, R.~N., {Hobbs}, G.~B., {Teoh}, A., \& {Hobbs}, M. 2005, \aj, 129,
  1993, \dodoi{10.1086/428488}

\bibitem[{{Mao} {et~al.}(2012){Mao}, {McClure-Griffiths}, {Gaensler},
  {Haverkorn}, {Beck}, {McConnell}, {Wolleben}, {Stanimirovi{\'c}}, {Dickey},
  \& {Staveley-Smith}}]{Mao2012}
{Mao}, S.~A., {McClure-Griffiths}, N.~M., {Gaensler}, B.~M., {et~al.} 2012,
  \apj, 759, 25, \dodoi{10.1088/0004-637X/759/1/25}

\bibitem[{{Mauch} {et~al.}(2003){Mauch}, {Murphy}, {Buttery}, {Curran},
  {Hunstead}, {Piestrzynski}, {Robertson}, \& {Sadler}}]{2003MNRAS.342.1117M}
{Mauch}, T., {Murphy}, T., {Buttery}, H.~J., {et~al.} 2003, \mnras, 342, 1117,
  \dodoi{10.1046/j.1365-8711.2003.06605.x}

\bibitem[{{McConnachie} {et~al.}(2005){McConnachie}, {Irwin}, {Ferguson},
  {Ibata}, {Lewis}, \& {Tanvir}}]{McConnachie2005}
{McConnachie}, A.~W., {Irwin}, M.~J., {Ferguson}, A.~M.~N., {et~al.} 2005,
  \mnras, 356, 979, \dodoi{10.1111/j.1365-2966.2004.08514.x}

\bibitem[{{McConnell} {et~al.}(1991){McConnell}, {McCulloch}, {Hamilton},
  {Ables}, {Hall}, {Jacka}, \& {Hunt}}]{1991MNRAS.249..654M}
{McConnell}, D., {McCulloch}, P.~M., {Hamilton}, P.~A., {et~al.} 1991, \mnras,
  249, 654, \dodoi{10.1093/mnras/249.4.654}

\bibitem[{{McConnell} {et~al.}(2020){McConnell}, {Hale}, {Lenc}, {Banfield},
  {Heald}, {Hotan}, {Leung}, {Moss}, {Murphy}, {O'Brien}, {Pritchard}, {Raja},
  {Sadler}, {Stewart}, {Thomson}, {Whiting}, {Allison}, {Amy}, {Anderson},
  {Ball}, {Bannister}, {Bell}, {Bock}, {Bolton}, {Bunton}, {Chippendale},
  {Collier}, {Cooray}, {Cornwell}, {Diamond}, {Edwards}, {Gupta}, {Hayman},
  {Heywood}, {Jackson}, {Koribalski}, {Lee-Waddell}, {McClure-Griffiths}, {Ng},
  {Norris}, {Phillips}, {Reynolds}, {Roxby}, {Schinckel}, {Shields},
  {Tremblay}, {Tzioumis}, {Voronkov}, \& {Westmeier}}]{2020PASA...37...48M}
{McConnell}, D., {Hale}, C.~L., {Lenc}, E., {et~al.} 2020, \pasa, 37, e048,
  \dodoi{10.1017/pasa.2020.41}

\bibitem[{{McCulloch} {et~al.}(1983){McCulloch}, {Hamilton}, {Ables}, \&
  {Hunt}}]{McCulloch1983}
{McCulloch}, P.~M., {Hamilton}, P.~A., {Ables}, J.~G., \& {Hunt}, A.~J. 1983,
  \nat, 303, 307, \dodoi{10.1038/303307a0}

\bibitem[{{McLaughlin} \& {Cordes}(2003)}]{McLaughlin2003}
{McLaughlin}, M.~A., \& {Cordes}, J.~M. 2003, \apj, 596, 982,
  \dodoi{10.1086/378232}

\bibitem[{{McMahon} {et~al.}(2013){McMahon}, {Banerji}, {Gonzalez}, {Koposov},
  {Bejar}, {Lodieu}, {Rebolo}, \& {VHS Collaboration}}]{McMahon2013}
{McMahon}, R.~G., {Banerji}, M., {Gonzalez}, E., {et~al.} 2013, The Messenger,
  154, 35

\bibitem[{{McMullin} {et~al.}(2007){McMullin}, {Waters}, {Schiebel}, {Young},
  \& {Golap}}]{2007ASPC..376..127M}
{McMullin}, J.~P., {Waters}, B., {Schiebel}, D., {Young}, W., \& {Golap}, K.
  2007, in Astronomical Society of the Pacific Conference Series, Vol. 376,
  Astronomical Data Analysis Software and Systems XVI, ed. R.~A. {Shaw},
  F.~{Hill}, \& D.~J. {Bell}, 127

\bibitem[{{Mikhailov} \& {van Leeuwen}(2016)}]{Mikhailov2016}
{Mikhailov}, K., \& {van Leeuwen}, J. 2016, \aap, 593, A21,
  \dodoi{10.1051/0004-6361/201628348}

\bibitem[{{Minniti} {et~al.}(2010){Minniti}, {Lucas}, {Emerson}, {Saito},
  {Hempel}, {Pietrukowicz}, {Ahumada}, {Alonso}, {Alonso-Garcia}, {Arias},
  {Bandyopadhyay}, {Barb{\'a}}, {Barbuy}, {Bedin}, {Bica}, {Borissova},
  {Bronfman}, {Carraro}, {Catelan}, {Clari{\'a}}, {Cross}, {de Grijs},
  {D{\'e}k{\'a}ny}, {Drew}, {Fari{\~n}a}, {Feinstein}, {Fern{\'a}ndez
  Laj{\'u}s}, {Gamen}, {Geisler}, {Gieren}, {Goldman}, {Gonzalez}, {Gunthardt},
  {Gurovich}, {Hambly}, {Irwin}, {Ivanov}, {Jord{\'a}n}, {Kerins}, {Kinemuchi},
  {Kurtev}, {L{\'o}pez-Corredoira}, {Maccarone}, {Masetti}, {Merlo},
  {Messineo}, {Mirabel}, {Monaco}, {Morelli}, {Padilla}, {Palma}, {Parisi},
  {Pignata}, {Rejkuba}, {Roman-Lopes}, {Sale}, {Schreiber}, {Schr{\"o}der},
  {Smith}, {}, {Soto}, {Tamura}, {Tappert}, {Thompson}, {Toledo}, {Zoccali}, \&
  {Pietrzynski}}]{Minniti2010}
{Minniti}, D., {Lucas}, P.~W., {Emerson}, J.~P., {et~al.} 2010, \na, 15, 433,
  \dodoi{10.1016/j.newast.2009.12.002}

\bibitem[{{Murphy} {et~al.}(2018){Murphy}, {Bolatto}, {Chatterjee}, {Casey},
  {Chomiuk}, {Dale}, {de Pater}, {Dickinson}, {Francesco}, {Hallinan},
  {Isella}, {Kohno}, {Kulkarni}, {Lang}, {Lazio}, {Leroy}, {Loinard},
  {Maccarone}, {Matthews}, {Osten}, {Reid}, {Riechers}, {Sakai}, {Walter}, \&
  {Wilner}}]{Murphy2018}
{Murphy}, E.~J., {Bolatto}, A., {Chatterjee}, S., {et~al.} 2018, in
  Astronomical Society of the Pacific Conference Series, Vol. 517, Science with
  a Next Generation Very Large Array, ed. E.~{Murphy}, 3.
\newblock \doarXiv{1810.07524}

\bibitem[{{Murphy} {et~al.}(2021){Murphy}, {Kaplan}, {Stewart}, {O'Brien},
  {Lenc}, {Pintaldi}, {Pritchard}, {Dobie}, {Fox}, {Leung}, {An}, {Bell},
  {Broderick}, {Chatterjee}, {Dai}, {d'Antonio}, {Doyle}, {Gaensler}, {Heald},
  {Horesh}, {Jones}, {McConnell}, {Moss}, {Raja}, {Ramsay}, {Ryder}, {Sadler},
  {Sivakoff}, {Wang}, {Wang}, {Wheatland}, {Whiting}, {Allison}, {Anderson},
  {Ball}, {Bannister}, {Bock}, {Bolton}, {Bunton}, {Chekkala}, {Chippendale},
  {Cooray}, {Gupta}, {Hayman}, {Jeganathan}, {Koribalski}, {Lee-Waddell},
  {Mahony}, {Marvil}, {McClure-Griffiths}, {Mirtschin}, {Ng}, {Pearce},
  {Phillips}, \& {Voronkov}}]{2021PASA...38...54M}
{Murphy}, T., {Kaplan}, D.~L., {Stewart}, A.~J., {et~al.} 2021, \pasa, 38,
  e054, \dodoi{10.1017/pasa.2021.44}

\bibitem[{{Norris} {et~al.}(2021){Norris}, {Marvil}, {Collier}, {Kapi{\'n}ska},
  {O'Brien}, {Rudnick}, {Andernach}, {Asorey}, {Brown}, {Br{\"u}ggen},
  {Crawford}, {English}, {Rahman}, {Filipovi{\'c}}, {Gordon}, {G{\"u}rkan},
  {Hale}, {Hopkins}, {Huynh}, {HyeongHan}, {James Jee}, {Koribalski}, {Lenc},
  {Luken}, {Parkinson}, {Prandoni}, {Raja}, {Reiprich}, {Riseley}, {Shabala},
  {Sheil}, {Vernstrom}, {Whiting}, {Allison}, {Anderson}, {Ball}, {Bell},
  {Bunton}, {Galvin}, {Gupta}, {Hotan}, {Jacka}, {Macgregor}, {Mahony}, {Maio},
  {Moss}, {Pandey-Pommier}, \& {Voronkov}}]{Norris2021}
{Norris}, R.~P., {Marvil}, J., {Collier}, J.~D., {et~al.} 2021, \pasa, 38,
  e046, \dodoi{10.1017/pasa.2021.42}

\bibitem[{{Ochsenbein} {et~al.}(2000){Ochsenbein}, {Bauer}, \&
  {Marcout}}]{2000A&AS..143...23O}
{Ochsenbein}, F., {Bauer}, P., \& {Marcout}, J. 2000, \aaps, 143, 23,
  \dodoi{10.1051/aas:2000169}

\bibitem[{{O'Dea}(1998)}]{O'Dea1998}
{O'Dea}, C.~P. 1998, \pasp, 110, 493, \dodoi{10.1086/316162}

\bibitem[{{Offringa} {et~al.}(2014){Offringa}, {McKinley}, {Hurley-Walker},
  {Briggs}, {Wayth}, {Kaplan}, {Bell}, {Feng}, {Neben}, {Hughes}, {Rhee},
  {Murphy}, {Bhat}, {Bernardi}, {Bowman}, {Cappallo}, {Corey}, {Deshpande},
  {Emrich}, {Ewall-Wice}, {Gaensler}, {Goeke}, {Greenhill}, {Hazelton},
  {Hindson}, {Johnston-Hollitt}, {Jacobs}, {Kasper}, {Kratzenberg}, {Lenc},
  {Lonsdale}, {Lynch}, {McWhirter}, {Mitchell}, {Morales}, {Morgan},
  {Kudryavtseva}, {Oberoi}, {Ord}, {Pindor}, {Procopio}, {Prabu}, {Riding},
  {Roshi}, {Shankar}, {Srivani}, {Subrahmanyan}, {Tingay}, {Waterson},
  {Webster}, {Whitney}, {Williams}, \& {Williams}}]{2014MNRAS.444..606O}
{Offringa}, A.~R., {McKinley}, B., {Hurley-Walker}, N., {et~al.} 2014, \mnras,
  444, 606, \dodoi{10.1093/mnras/stu1368}

\bibitem[{{Park} {et~al.}(2012){Park}, {Hughes}, {Slane}, {Burrows}, {Lee}, \&
  {Mori}}]{2012ApJ...748..117P}
{Park}, S., {Hughes}, J.~P., {Slane}, P.~O., {et~al.} 2012, \apj, 748, 117,
  \dodoi{10.1088/0004-637X/748/2/117}

\bibitem[{{Pennock} {et~al.}(2021){Pennock}, {van Loon}, {Filipovi{\'c}},
  {Andernach}, {Haberl}, {Kothes}, {Lenc}, {Rudnick}, {White}, {Agliozzo},
  {Ant{\'o}n}, {Boji{\v{c}}i{\'c}}, {Bomans}, {Collier}, {Crawford}, {Hopkins},
  {Jeganathan}, {Kavanagh}, {Koribalski}, {Leahy}, {Maggi}, {Maitra}, {Marvil},
  {Micha{\l}owski}, {Norris}, {Oliveira}, {Payne}, {Sano}, {Sasaki},
  {Staveley-Smith}, \& {Vardoulaki}}]{2021MNRAS.506.3540P}
{Pennock}, C.~M., {van Loon}, J.~T., {Filipovi{\'c}}, M.~D., {et~al.} 2021,
  \mnras, 506, 3540, \dodoi{10.1093/mnras/stab1858}

\bibitem[{{Pietrzy{\'n}ski} {et~al.}(2013){Pietrzy{\'n}ski}, {Graczyk},
  {Gieren}, {Thompson}, {Pilecki}, {Udalski}, {Soszy{\'n}ski}, {Koz{\l}owski},
  {Konorski}, {Suchomska}, {Bono}, {Moroni}, {Villanova}, {Nardetto},
  {Bresolin}, {Kudritzki}, {Storm}, {Gallenne}, {Smolec}, {Minniti}, {Kubiak},
  {Szyma{\'n}ski}, {Poleski}, {Wyrzykowski}, {Ulaczyk}, {Pietrukowicz},
  {G{\'o}rski}, \& {Karczmarek}}]{Pietrzynski2013}
{Pietrzy{\'n}ski}, G., {Graczyk}, D., {Gieren}, W., {et~al.} 2013, \nat, 495,
  76, \dodoi{10.1038/nature11878}

\bibitem[{{Pintaldi} {et~al.}(2021){Pintaldi}, {Stewart}, {O'Brien}, {Kaplan},
  \& {Murphy}}]{2021arXiv210105898P}
{Pintaldi}, S., {Stewart}, A., {O'Brien}, A., {Kaplan}, D., \& {Murphy}, T.
  2021, arXiv e-prints, arXiv:2101.05898.
\newblock \doarXiv{2101.05898}

\bibitem[{{Pritchard} {et~al.}(2021){Pritchard}, {Murphy}, {Zic}, {Lynch},
  {Heald}, {Kaplan}, {Anderson}, {Banfield}, {Hale}, {Hotan}, {Lenc}, {Leung},
  {McConnell}, {Moss}, {Raja}, {Stewart}, \& {Whiting}}]{2021MNRAS.502.5438P}
{Pritchard}, J., {Murphy}, T., {Zic}, A., {et~al.} 2021, \mnras, 502, 5438,
  \dodoi{10.1093/mnras/stab299}

\bibitem[{{Purcell} {et~al.}(2020){Purcell}, {Van Eck}, {West}, {Sun}, \&
  {Gaensler}}]{Purcell2020}
{Purcell}, C.~R., {Van Eck}, C.~L., {West}, J., {Sun}, X.~H., \& {Gaensler},
  B.~M. 2020, {RM-Tools: Rotation measure (RM) synthesis and Stokes
  QU-fitting}.
\newblock \doeprint{2005.003}

\bibitem[{{Radhakrishnan} \& {Cooke}(1969)}]{Radhakrishnan1969}
{Radhakrishnan}, V., \& {Cooke}, D.~J. 1969, \aplett, 3, 225

\bibitem[{{Ransom}(2011)}]{Ransom2011}
{Ransom}, S. 2011, {PRESTO: PulsaR Exploration and Search TOolkit}.
\newblock \doeprint{1107.017}

\bibitem[{{Ransom}(2001)}]{2001PhDT.......123R}
{Ransom}, S.~M. 2001, PhD thesis, Harvard University

\bibitem[{{Ransom} {et~al.}(2003){Ransom}, {Cordes}, \&
  {Eikenberry}}]{Ransom2003}
{Ransom}, S.~M., {Cordes}, J.~M., \& {Eikenberry}, S.~S. 2003, \apj, 589, 911,
  \dodoi{10.1086/374806}

\bibitem[{{Ransom} {et~al.}(2002){Ransom}, {Eikenberry}, \&
  {Middleditch}}]{Ransom+2002}
{Ransom}, S.~M., {Eikenberry}, S.~S., \& {Middleditch}, J. 2002, \aj, 124,
  1788, \dodoi{10.1086/342285}

\bibitem[{{Rickett}(1990)}]{1990ARA&A..28..561R}
{Rickett}, B.~J. 1990, \araa, 28, 561,
  \dodoi{10.1146/annurev.aa.28.090190.003021}

\bibitem[{{Ridley} {et~al.}(2013){Ridley}, {Crawford}, {Lorimer}, {Bailey},
  {Madden}, {Anella}, \& {Chennamangalam}}]{Ridley2013}
{Ridley}, J.~P., {Crawford}, F., {Lorimer}, D.~R., {et~al.} 2013, \mnras, 433,
  138, \dodoi{10.1093/mnras/stt709}

\bibitem[{{Ridolfi} {et~al.}(2021){Ridolfi}, {Gautam}, {Freire}, {Ransom},
  {Buchner}, {Possenti}, {Venkatraman Krishnan}, {Bailes}, {Kramer},
  {Stappers}, {Abbate}, {Barr}, {Burgay}, {Camilo}, {Corongiu}, {Jameson},
  {Padmanabh}, {Vleeschower}, {Champion}, {Chen}, {Geyer}, {Karastergiou},
  {Karuppusamy}, {Parthasarathy}, {Reardon}, {Serylak}, {Shannon}, \&
  {Spiewak}}]{Ridolfi+2021}
{Ridolfi}, A., {Gautam}, T., {Freire}, P.~C.~C., {et~al.} 2021, \mnras, 504,
  1407, \dodoi{10.1093/mnras/stab790}

\bibitem[{{Robitaille} \& {Bressert}(2012)}]{Robitaille2012}
{Robitaille}, T., \& {Bressert}, E. 2012, {APLpy: Astronomical Plotting Library
  in Python}.
\newblock \doeprint{1208.017}

\bibitem[{{Rowlinson} {et~al.}(2019){Rowlinson}, {Stewart}, {Broderick},
  {Swinbank}, {Wijers}, {Carbone}, {Cendes}, {Fender}, {van der Horst},
  {Molenaar}, {Scheers}, {Staley}, {Farrell}, {Grie{\ss}meier}, {Bell},
  {Eisl{\"o}ffel}, {Law}, {van Leeuwen}, \& {Zarka}}]{2019A&C....27..111R}
{Rowlinson}, A., {Stewart}, A.~J., {Broderick}, J.~W., {et~al.} 2019, Astronomy
  and Computing, 27, 111, \dodoi{10.1016/j.ascom.2019.03.003}

\bibitem[{{Rubio-Herrera} {et~al.}(2013){Rubio-Herrera}, {Stappers}, {Hessels},
  \& {Braun}}]{Rubio-Herrera2013}
{Rubio-Herrera}, E., {Stappers}, B.~W., {Hessels}, J.~W.~T., \& {Braun}, R.
  2013, \mnras, 428, 2857, \dodoi{10.1093/mnras/sts205}

\bibitem[{{Sault} {et~al.}(1995){Sault}, {Teuben}, \&
  {Wright}}]{1995ASPC...77..433S}
{Sault}, R.~J., {Teuben}, P.~J., \& {Wright}, M.~C.~H. 1995, in Astronomical
  Society of the Pacific Conference Series, Vol.~77, Astronomical Data Analysis
  Software and Systems IV, ed. R.~A. {Shaw}, H.~E. {Payne}, \& J.~J.~E.
  {Hayes}, 433.
\newblock \doarXiv{astro-ph/0612759}

\bibitem[{{Shimwell} {et~al.}(2017){Shimwell}, {R{\"o}ttgering}, {Best},
  {Williams}, {Dijkema}, {de Gasperin}, {Hardcastle}, {Heald}, {Hoang},
  {Horneffer}, {Intema}, {Mahony}, {Mandal}, {Mechev}, {Morabito}, {Oonk},
  {Rafferty}, {Retana-Montenegro}, {Sabater}, {Tasse}, {van Weeren},
  {Br{\"u}ggen}, {Brunetti}, {Chy{\.z}y}, {Conway}, {Haverkorn}, {Jackson},
  {Jarvis}, {McKean}, {Miley}, {Morganti}, {White}, {Wise}, {van Bemmel},
  {Beck}, {Brienza}, {Bonafede}, {Calistro Rivera}, {Cassano}, {Clarke},
  {Cseh}, {Deller}, {Drabent}, {van Driel}, {Engels}, {Falcke}, {Ferrari},
  {Fr{\"o}hlich}, {Garrett}, {Harwood}, {Heesen}, {Hoeft}, {Horellou},
  {Israel}, {Kapi{\'n}ska}, {Kunert-Bajraszewska}, {McKay}, {Mohan},
  {Orr{\'u}}, {Pizzo}, {Prandoni}, {Schwarz}, {Shulevski}, {Sipior}, {Smith},
  {Sridhar}, {Steinmetz}, {Stroe}, {Varenius}, {van der Werf}, {Zensus}, \&
  {Zwart}}]{Shimwell2017}
{Shimwell}, T.~W., {R{\"o}ttgering}, H.~J.~A., {Best}, P.~N., {et~al.} 2017,
  \aap, 598, A104, \dodoi{10.1051/0004-6361/201629313}

\bibitem[{{Skrutskie} {et~al.}(2006){Skrutskie}, {Cutri}, {Stiening},
  {Weinberg}, {Schneider}, {Carpenter}, {Beichman}, {Capps}, {Chester},
  {Elias}, {Huchra}, {Liebert}, {Lonsdale}, {Monet}, {Price}, {Seitzer},
  {Jarrett}, {Kirkpatrick}, {Gizis}, {Howard}, {Evans}, {Fowler}, {Fullmer},
  {Hurt}, {Light}, {Kopan}, {Marsh}, {McCallon}, {Tam}, {Van Dyk}, \&
  {Wheelock}}]{2006AJ....131.1163S}
{Skrutskie}, M.~F., {Cutri}, R.~M., {Stiening}, R., {et~al.} 2006, \aj, 131,
  1163, \dodoi{10.1086/498708}

\bibitem[{{Smith} \& {MCELS Team}(1998)}]{Smith1998}
{Smith}, R.~C., \& {MCELS Team}. 1998, \pasa, 15, 163, \dodoi{10.1071/AS98163}

\bibitem[{{Strom}(1987)}]{1987ApJ...319L.103S}
{Strom}, R.~G. 1987, \apjl, 319, L103, \dodoi{10.1086/184963}

\bibitem[{{Taylor} \& {Cordes}(1993)}]{Taylor1993}
{Taylor}, J.~H., \& {Cordes}, J.~M. 1993, \apj, 411, 674,
  \dodoi{10.1086/172870}

\bibitem[{{Tuntsov} {et~al.}(2017){Tuntsov}, {Stevens}, {Bannister}, {Bignall},
  {Johnston}, {Reynolds}, \& {Walker}}]{2017MNRAS.469.5023T}
{Tuntsov}, A.~V., {Stevens}, J., {Bannister}, K.~W., {et~al.} 2017, \mnras,
  469, 5023, \dodoi{10.1093/mnras/stx1223}

\bibitem[{{van Leeuwen} {et~al.}(2020){van Leeuwen}, {Mikhailov}, {Keane},
  {Coenen}, {Connor}, {Kondratiev}, {Michilli}, \& {Sanidas}}]{vanLeeuwen2020}
{van Leeuwen}, J., {Mikhailov}, K., {Keane}, E., {et~al.} 2020, \aap, 634, A3,
  \dodoi{10.1051/0004-6361/201937065}

\bibitem[{{Vedantham} {et~al.}(2020){Vedantham}, {Callingham}, {Shimwell},
  {Dupuy}, {Best}, {Liu}, {Zhang}, {De}, {Lamy}, {Zarka}, {R{\"o}ttgering}, \&
  {Shulevski}}]{2020ApJ...903L..33V}
{Vedantham}, H.~K., {Callingham}, J.~R., {Shimwell}, T.~W., {et~al.} 2020,
  \apjl, 903, L33, \dodoi{10.3847/2041-8213/abc256}

\bibitem[{{Walker}(1998)}]{1998MNRAS.294..307W}
{Walker}, M.~A. 1998, \mnras, 294, 307,
  \dodoi{10.1046/j.1365-8711.1998.01238.x}

\bibitem[{{Wang} {et~al.}(2021){Wang}, {Kaplan}, {Murphy}, {Lenc}, {Dai},
  {Barr}, {Dobie}, {Gaensler}, {Heald}, {Leung}, {O'Brien}, {Pintaldi},
  {Pritchard}, {Rea}, {Sivakoff}, {Stappers}, {Stewart}, {Tremou}, {Wang},
  {Woudt}, \& {Zic}}]{2021ApJ...920...45W}
{Wang}, Z., {Kaplan}, D.~L., {Murphy}, T., {et~al.} 2021, \apj, 920, 45,
  \dodoi{10.3847/1538-4357/ac2360}

\bibitem[{{Wright} {et~al.}(2010){Wright}, {Eisenhardt}, {Mainzer}, {Ressler},
  {Cutri}, {Jarrett}, {Kirkpatrick}, {Padgett}, {McMillan}, {Skrutskie},
  {Stanford}, {Cohen}, {Walker}, {Mather}, {Leisawitz}, {Gautier}, {McLean},
  {Benford}, {Lonsdale}, {Blain}, {Mendez}, {Irace}, {Duval}, {Liu}, {Royer},
  {Heinrichsen}, {Howard}, {Shannon}, {Kendall}, {Walsh}, {Larsen}, {Cardon},
  {Schick}, {Schwalm}, {Abid}, {Fabinsky}, {Naes}, \&
  {Tsai}}]{2010AJ....140.1868W}
{Wright}, E.~L., {Eisenhardt}, P. R.~M., {Mainzer}, A.~K., {et~al.} 2010, \aj,
  140, 1868, \dodoi{10.1088/0004-6256/140/6/1868}

\bibitem[{{Yao} {et~al.}(2017){Yao}, {Manchester}, \&
  {Wang}}]{2017ApJ...835...29Y}
{Yao}, J.~M., {Manchester}, R.~N., \& {Wang}, N. 2017, \apj, 835, 29,
  \dodoi{10.3847/1538-4357/835/1/29}

\bibitem[{{Young} {et~al.}(2010){Young}, {Chan}, {Burman}, \&
  {Blair}}]{2010MNRAS.402.1317Y}
{Young}, M.~D.~T., {Chan}, L.~S., {Burman}, R.~R., \& {Blair}, D.~G. 2010,
  \mnras, 402, 1317, \dodoi{10.1111/j.1365-2966.2009.15972.x}

\bibitem[{{Zaritsky} {et~al.}(2004){Zaritsky}, {Harris}, {Thompson}, \&
  {Grebel}}]{2004AJ....128.1606Z}
{Zaritsky}, D., {Harris}, J., {Thompson}, I.~B., \& {Grebel}, E.~K. 2004, \aj,
  128, 1606, \dodoi{10.1086/423910}

\end{thebibliography}
\bibliographystyle{aasjournal}


\listofchanges

\end{document}